# The Human Genomic Landscape of Oceania

**Consuelo D. Quinto-Cortés*[1], Carmina Barberena-Jonas*[1], Peter A. Gerlach*[2], Sofía Vieyra-Sánchez[1], Ram González-Buenfil[1], Petria R. Russell[3], Stephen J. Oppenheimer[4], Kathryn Auckland[5], Kathryn Robson[4,27], Tom Parks[4], Julian R. Homburger[2], Alissa L. Severson[2], Genevieve L. Wojcik[6], J. Víctor Moreno-Mayar[7], Javier Blanco-Portillo[8], Karla Sandoval[1], Ishwar Verma[9,27], Abdul Salam M. Sofro[9], Frederick Delfin[10], Maria Corazon A. De Ungria[10], George Koki[11], George Aho[12], Jonathan S. Friedlaender[13], Francoise Friedlaender[13,27], Angela Allen[14,15], Stephen Allen[15,16], Adrian V. S. Hill[4], Toomas Kivisild[17], Phillip Endicott[18,19,20,21], Matthew Spriggs[22,23], Mark Stoneking[24,25], Carlos D. Bustamante[2], Maude Phipps[26], William Pomat[11,**], Alexander J. Mentzer[4,**], Andrés Moreno-Estrada[1,**], and Alexander G. Ioannidis[2,3,**]**

[1]National Laboratory of Genomics for Biodiversity (LANGEBIO)—Advanced Genomics Unit, CINVESTAV, Mexico
[2]Department of Genetics, Stanford University, CA, USA
[3]Genomics Institute, University of California Santa Cruz, CA, USA
[4]University of Oxford, United Kingdom
[5]Department of Medicine, University of Cambridge, United Kingdom
[6]Johns Hopkins Bloomberg School of Public Health, USA
[7]Centre for GeoGenetics, University of Copenhagen, Denmark
[8]Department for Biology, Stanford University, Stanford, CA, USA
[9]Department of Biochemistry, Faculty of Medicine, Yayasan Rumah Sakit Islam (YARSI) University, Cempaka Putih, Jakarta, Indonesia
[10]DNA Analysis Laboratory, Natural Sciences Research Institute, University of the Philippines Diliman, Philippines
[11]Papua New Guinea Institute of Medical Research, Papua New Guinea
[12]Vaiola Hospital, Tonga
[13]Department of Anthropology, Temple University, USA
[14]MRC Weatherall Institute of Molecular Medicine, Headley Way, Headington, Oxford, OX3 9DS, UK
[15]Edward Francis Small Teaching Hospital, Republic of The Gambia
[16]Liverpool School of Tropical Medicine, UK
[17]Katholieke Universiteit, Belgium
[18]University of Tartu, Institute of Genomics, 23b Riia, Tartu, 51010, Estonia
[19]Department of Archaeology and Anthropology, Bournemouth University, Fern Barrow, Poole, Dorset, BH12 5BB, UK
[20]Department of Linguistics, University of Hawai'i at Mānoa, Honolulu, Hawai'i, 96822, USA
[21]DFG Center for Advanced Studies, University of Tübingen, Tübingen, 72074, Germany
[22]The Australian National University, Australia
[23]Vanuatu Cultural Centre/Kaljoral Senta, Port Vila, Vanuatu
[24]Department of Evolutionary Genetics, Max Planck Institute for Evolutionary Anthropology, Germany
[25]Laboratoire Biométrie et Biologie Évolutive, UMR 5558, CNRS & Université de Lyon, Lyon, France
[26]Jeffrey Cheah School of Medicine and Health Sciences, Monash University Malaysia, Malaysia
[27]Deceased
*These authors contributed equally to this work
**Correspondence: william.pomat@pngimr.org.pg, alexander.mentzer@ndm.ox.ac.uk, andres.moreno@cinvestav.mx, ioannidis@stanford.edu

**Abstract**

Oceania and Island Southeast Asia have a rich, yet understudied, human genomic landscape. This region encompasses some of the first areas inhabited by humans following the out-of-Africa expansion, includes populations with the highest levels of archaic hominin introgression, and contains Pacific islands that are among the most remote continuously inhabited locations in the world. Here, we describe the first region-wide analysis of individuals from population groups spanning Oceania and its broad perimeter. In total we generate and analyze genome-wide data from 92 different populations, 58 separate islands, and 30 countries, covering one third of the planet. Leveraging this diverse dataset, we resolve genetic connections among islands, providing a detailed view of regional population structure and identifying the island groups involved in the settlement of several Polynesian Outliers. Ancestry-specific analyses allow us to deconvolve different layers of history, from tracing groups deriving their Austronesian ancestry via the Lapita expansion to quantifying variable archaic introgression across the basal Papuan component of Oceanians and Southeast Asians. Finally, we map biomedically relevant variants across Oceania and Southeast Asia, observing pronounced allele-frequency differences between population groups. Together, these findings refine models of oceanic settlement and admixture and establish a comprehensive reference that will advance global efforts to ensure broad and equitable representation in human genomics.

## Introduction

Oceania and its surroundings comprise island and coastal populations that have histories of settlement dating back at least 50,000 years with some of the earliest evidence found in the highlands of New Guinea[1]. Later migrations across Oceania continued into relatively recent times, from the Lapita expansion (3200–2800 BP) reaching Tonga and Samoa, to the settlement of eastern Polynesia (1200–700 BP)[2,3]. The region's extraordinary cultural, sociopolitical, and linguistic diversity reflects complex population histories shaped by periods of isolation as well as moments of contact and migration across vast ocean tracts. Although several populations from Oceania have previously been studied using classical and uniparental markers[4–7] and more recently, a few sub-regions have been investigated using genome-wide and whole genome data[8–11], there is still a large gap in our knowledge of pan-Oceanian connections, and the genetic diversity of this space as a whole.

The Oceania Genome Variation Project (OGVP) represents the most comprehensive effort to date to survey human genetic diversity across Oceania, Island and Peninsular Southeast Asia, and the Indian Ocean islands, generating and analyzing genome-wide data from 92 populations across 58 islands (Figure 1A). This dataset spans the full breadth of Oceania and its periphery, including groups from Micronesia, Melanesia, and Polynesia as well as from mainland and island Southeast Asia, the Andaman Islands, and Taiwan, locations with deep historical and linguistic ties to the Pacific. To provide a comprehensive perspective on Austronesian dispersal and admixture, we also include several groups from the island of Madagascar, the westernmost extent of the Austronesian expansion. This resource is intended to support future research into the genetic and biomedically related variation of Oceania and Southeast Asia, while contributing to broader global efforts toward equitable genomic and precision health.

## Results

### Genetic variation in Oceania and its periphery

To explore the genetic relationships between populations, we investigate unsupervised genetic clustering using ADMIXTURE[12]. At K = 9 (Figure 1B; full range K = 2–17 in Figures S1–S3), we identify region-enriched components corresponding to Papuan (Clusters 6 and 8), Micronesian (Cluster 7), and Polynesian (Cluster 4) regions, reflecting broad cultural and linguistic differences between these areas[13–16]. The Papuan-like clusters (6 and 8) further separate into New Guinea– and New Britain–associated components, respectively. An Island East Asian–like component (Cluster 5) forms a west-to-east cline that could be a combination of both the legacy of the Austronesian expansion and modern admixture events. This component is also present at high proportions in Madagascar, together with an African-like component (Cluster 2), consistent with previous studies and the Austronesian languages of Madagascar[17]. A European-like component (Cluster 9) is present in several groups across Oceania and Island and Peninsular Southeast Asia, indicating more recent colonial admixture. These broad regional patterns are also reflected in genome-wide differentiation ($F_{ST}$; Fig. S4) and in uniparental markers of mtDNA and Y-chromosome (Figures S5 and S6). While the clustering patterns align in part with the regional labels of Melanesia (Papuan component), Micronesia, and Polynesia, we recognize that these terms have origins in early European attempts to classify Pacific Islands and have been critiqued for oversimplifying Oceania[18–20]. Indeed, as we will show, there are numerous exceptions and complexities to these regional categories; however, we use them as geographic reference points, while recognizing their limitations.

Despite the clusters broadly reflecting regional variation, there are also strikingly distinct stories embedded in them. In Polynesia we observe a clear difference in the ancestry components of Tonga and Samoa (westernmost Polynesian islands) compared to the rest of Polynesia. This pattern likely indicates the recent founder effects that occurred during the settlement of many of these islands, with the easternmost Polynesian islands being more directly related to each other through a common founder effect, anchoring this clustering component. This is consistent with the hypothesis that the historical homeland of the Polynesian cultures was located in the region of Tonga and Samoa from which individuals migrated out to the more eastern Polynesian islands in the last two millennia[2,21]. A similar pattern is observed within fourteen Tuvaluan individuals, who have ancestry proportions similar to those in Tonga and Samoa (Figure S1 and Figure S7), consistent with possible settlement of Tuvalu by western Polynesians from Samoa and/or Tonga[22]. In contrast, five Tuvaluan individuals show ancestry components more similar to individuals from Micronesia (Figure S7). Tuvalu is the only population in our dataset with a pronounced within-population split in ancestry components. All other population groups show individuals with consistent cluster compositions. This apparent bifurcation in Tuvalu may reflect structure amongst that nation's islands. While a Polynesian language is spoken on most islands in Tuvalu, a Micronesian language is spoken on the island of Nui. In subsequent analyses, we refer to these two different clusters as "Tuvalu Polynesia" and "Tuvalu Micronesia" to differentiate the Polynesian-like and Micronesian-like individuals.

Other notable stories that deviate from broader regional patterns include evidence of Micronesian-related ancestry in Palau and Guam that can now be resolved more clearly in a region-wide framework, as well as Polynesian ancestry in populations along the southern coast of New Guinea (Supplementary Note 2). Together, these findings highlight the value of region-wide sampling for resolving complex population histories across Oceania.

### Ancestry specific analyses deconvolve separate Austronesian and Papuan histories

As the Lapita cultural complex played a key role in the early settlement of Remote Oceania, we incorporated eight published ancient individuals from Lapita sites in Vanuatu and Tonga[23,24] into our PCA (Figures 2A; for PCA with Human Genome Diversity Project (HGDP) reference populations see S8). In the PCA, these ancient individuals cluster distinctly and do not overlap with present-day Oceanian groups. This pattern is consistent with prior findings showing that Lapita derived ancestry is not present in an unadmixed form among present-day Oceanian groups[23–25].

To situate the ancient Lapita individuals relative to present-day populations, we performed local ancestry inference and partitioned each present-day genome into European-, Papuan-, Austronesian-, and African-like segments (Figures 2B and S3). Previous studies have shown that Pacific Islanders derive much of their ancestry from two different ancestral groups: a source related to early settlers of Island Southeast Asia $\sim$50,000 years ago (here termed "Papuan" reflecting the largest island that is dominated by it) and a more recent ancestry related to indigenous Taiwanese groups (here termed "Austronesian" reflecting the associated language family)[23,26]. Using segments assigned to the Papuan and Austronesian ancestries, we computed pairwise distances and constructed neighbor-joining trees for each ancestry component in each present-day group, rooted with the Andamanese population as an outgroup (Figures 2C and S9).

In the Austronesian component tree, Banks & Torres (Vanuatu) root the clade containing Fiji, Polynesian groups, and Polynesian outliers (Figure 2C). The ancient Lapita individuals branch even earlier, basal to this entire grouping as well as to the Micronesian grouping; the latter is consistent with the findings of Liu et al. 2022 on the origin of Micronesians[27]. This ancestry-specific Austronesian tree structure is consistent with Lapita-related groups being ancestral to all present-day Polynesian and Central/Eastern Micronesian groups' Austronesian component, with Banks & Torres and Fiji occupying genetic and geographic threshold positions with respect to the Polynesian clade. Furthermore, basal to all Oceanian groups are indigenous Taiwanese populations, consistent with the deep ancestral origin of the Lapita complex in Austronesian ancestry originating ultimately in ancient Taiwan (Figure S9).

It is noteworthy that the Austronesian ancestry-specific tree shows present-day populations of Vanuatu branching both earlier (Port Orly and Maewo) and later (Banks & Torres) than the ancient Lapita individuals (Figure S9). That the branching spectrum of present-day Austronesian ancestry across the islands of Vanuatu encompasses the ancient Lapita, suggests that the Austronesian component in the ancient Lapita individuals are part of the ancestral Vanuatu variation, rather than representing a single relic or reintroduced lineage from outside it. Furthermore, it suggests that basal ancestral Austronesian variation amongst the islands of Vanuatu associated with the ancient Lapita is preserved in the present-day Ni-Vanuatu, consistent with long-term linguistic and genealogical continuity in the native people of Vanuatu[24]. In contrast, the findings are less compatible with the hypothesis of a near-complete population turnover in Vanuatu, in which Lapita-associated Austronesian ancestry was for a time lost and later re-introduced[25].

In the Papuan-component tree Banks & Torres (Vanuatu) clusters with nearby Island Melanesian groups (including Fiji) but also clusters with Papuan-speaking groups from New Britain, with no close clustering to Polynesian groups. This pattern is consistent with prior work showing that, among Near Oceanian populations, Vanuatu's Papuan-ancestry profile is most similar to groups from New Britain in the Bismarck Archipelago, supporting a Bismarck-derived contribution to Papuan ancestry in Vanuatu, rather than a source in the geographically closer Solomon Islands that lie in between the Bismarks and Vanuatu[25].

In summary, the different topology of Banks & Torres between the Austronesian and Papuan ancestry-specific trees reflects the very different migration histories of those two ancestries and illustrates the power and importance of performing ancestry-specific analyses, rather than using per-individual or per-population averaged statistics, in such admixed scenarios. More broadly, our results align with the prevailing model in which the Lapita people led the initial settlement into western Polynesia via Island Melanesia, followed later by expansions into eastern Polynesia and Micronesia. However, in contrast to a previous model proposing major population replacement of Austronesian ancestry, our findings are consistent with the Ni-Vanuatu retaining substantial basal, Lapita-related Austronesian ancestry variation[24], rather than experiencing near complete replacement of this component[25].

## Polynesian Outliers

Polynesia is often defined geographically as the triangle linking Hawaiʻi, Aotearoa New Zealand, and Rapa Nui. Yet a set of islands known as the Polynesian Outliers lie well beyond this core region, scattered westward across Island Melanesia and Micronesia. These outlier communities speak Polynesian languages and maintain many cultural traditions associated with societies inside the Polynesian triangle[28,29]. Linguistic, ethnographic, and archaeological evidence consistently indicates that the Polynesian Outliers derive from West Polynesian populations and are closely related to populations within the Polynesian triangle[30–35]. However, the later role of the Polynesian Outliers in the settlement of Eastern Polynesia remains debated: while many studies link the settlement of Eastern Polynesia to Western Polynesia[29,36], other studies have argued for the settlement of Eastern Polynesia from the North-Central Polynesian Outliers of the Solomon Islands, of which Ontong Java is the largest[37–39]. To date, genome-wide genetic studies including Polynesian Outlier communities have been few and often sidelined within broader surveys of Pacific variation, leaving the population history of the Outliers themselves much less well characterized than that of groups inside the Polynesian triangle[23–25,27,40,41].

To better understand the population history of these communities, we examine the genetic relationships between Polynesian Outliers in the OGVP cohort—Kapingamarangi, Ontong Java, Tikopia, and Rennell–Bellona—and populations in Micronesia and in West and Eastern Polynesia. In unsupervised genetic clustering we find that Ontong Java, Tikopia, and Rennell–Bellona, although located geographically in Melanesia, carry a much larger proportion of their ancestry from a Polynesian-like ancestry component (Cluster 4 in Figure 1B) compared to their neighboring islands, consistent with their classification as Polynesian Outliers. Additionally, these communities also harbor substantial Micronesian-like genetic ancestry (Cluster 7 in Figure 1B), suggesting they experienced additional gene flow from Micronesian populations. By comparison, Kapingamarangi located in Micronesia carries a smaller Polynesian-like component than Ontong Java, Tikopia, and Rennell–Bellona, with the majority of its ancestry coming from the Micronesian-like component.

To directly investigate the origins of migration, we calculated the total amount of shared identity-by-descent (IBD) between populations in the OGVP cohort, normalized by the number of individual pairs (Figure 3A; for within region ROH see Figures S10-S12). We first analyzed the Polynesian Outliers jointly as a single group (Figure 3A). Here, we observe the strongest IBD sharing between the Polynesian Outliers and Tuvalu in West Polynesia and Kiribati in Micronesia. Notably, the Micronesian-like ancestry in the Polynesian Outliers and their high IBD sharing with Micronesian islands are consistent with documented cultural contacts between Micronesia and Melanesia. These contacts include shared practices such as backstrap loom weaving, kite fishing, and shell artefact traditions, which have been suggested to have spread from Micronesia via the Polynesian Outliers as far south as the Banks and Torres islands and Santo in Vanuatu[42].

We next examined the IBD between the Polynesian Outliers and Polynesia and Micronesia separately to resolve population-specific patterns (Figure 3B). Among the Outliers, Ontong Java is especially informative of the role of Polynesian Outliers in the settlement of Eastern Polynesia, as the North-Central Outlier model links the settlement of Eastern Polynesia to Northern Outliers including Ontong Java, rather than Western Polynesia[37,38]. Thus, under the North-Central Outlier model, Ontong Java would be expected to share more IBD with Eastern Polynesia than with Western Polynesia, because a contribution from Ontong Java to Eastern Polynesia would represent a more recent shared history than the earlier settlement of Ontong Java from Western Polynesia, and IBD sharing is most sensitive to recent shared ancestry. Instead, for Ontong Java we observe the highest IBD sharing with Tuvalu among Polynesian groups, indicating that Ontong Java's Polynesian ancestry is more closely tied to Western Polynesia than to Eastern Polynesia and therefore supporting a primary settlement of Eastern Polynesia via Western Polynesia rather than from Northern Outliers[29,36]. For Micronesian groups we observe the highest IBD sharing with Kiribati. Together, this pattern is consistent with an earlier lexicostatistical reconstruction which proposed multiple episodes of settlement of Ontong Java from Tuvaluan-related Polynesian groups and secondary contact with Micronesian groups primarily from Kiribati[32].

Similarly, Tikopia shows high IBD sharing with Tuvalu amongst Polynesian populations and with Kiribati amongst Micronesian populations, but additionally exhibits high Polynesian IBD sharing with Samoa and Mauke, suggesting broader connections within West and Central Polynesia. The elevated IBD of Tikopia with Tuvalu and Samoa aligns with oral-historical and ethnographic evidence for Western Polynesian origin[43]. Although prior work has not documented admixture between Tikopia and Kiribati, our IBD analysis suggests that Tikopia has also received substantial Micronesian gene flow from Kiribati and possibly from additional Micronesian groups, resulting in the Micronesian-like ancestry of Tikopia observed in Figure 1B. For Rennell and Bellona, we observe elevated IBD sharing with Tuvalu and Samoa amongst Polynesian references, but no single Micronesian source clearly accounts for its Micronesian-related ancestry. This is consistent with with oral-historical evidence suggesting links to Western Polynesia[44]. Similar to our IBD analysis, prior work has not identified a single clear source for the Micronesian-like ancestry in Rennell and Bellona, which might suggest that this ancestry is attributable to broader, low-level contact with Micronesian groups. Finally, for Kapingamarangi we again observe highest IBD to Tuvalu among Polynesian populations and Kiribati amongst Micronesian populations. This is consistent with linguistic evidence grouping the language of Kapingamarangi with Tuvaluan[36] and folklore on Kapingamarangi describing a voyage from Tamana Island in Kiribati to Kapingamarangi[45]. Although each outlier shows affinities to West Polynesian and Micronesian sources, the Polynesian Outliers share only modest amounts of IBD with one another (Figure 3C), suggesting limited recent shared ancestry among outlier populations and supporting largely independent demographic histories after settlement. The same broad pattern is observed with outgroup $f_3$, albeit with less resolution for island-specific relationships (Figure S13).

In summary, our analyses support a model in which Polynesian Outliers were founded through back-migration from West Polynesia, most likely involving Tuvalu, while also receiving substantial Micronesian-related ancestry from Kiribati. Furthermore, our results strongly support the model in which Eastern Polynesia was settled primarily by Western Polynesia rather than from Northern Polynesia Outliers.

## Denisovan introgression in the Pacific

Oceanian populations harbor the highest levels of Denisovan ancestry globally[46,47]. To quantify the Denisovan ancestry for populations in OGVP, we ran an F4-ratio[48] test using an established test topology meant to isolate Denisovan signals unique to Oceanians by using East Asians to control for possible confounding Neanderthal ancestry[49]. The ratio tested was f4(Altai Neanderthal, Yoruba; OGVP, Han)/f4(Altai Neanderthal, Yoruba; Denisovan, Han). We applied this to three sets of our data: (i) global data (non-ancestry-specific), (ii) Papuan component, and (iii) Austronesian component segments from our local ancestry inference (Figure 4). Our results on global ancestry show an average of 3.2% Denisovan introgression in Papuan Highlanders, with the second-highest values found in native northern Philippines populations (Aeta and Agta) (Figure 4A), consistent with prior studies finding the highest Denisovan ancestry in Papuans followed by Philippine Negrito groups[47,49–52]. We also identify low but detectable signals of introgression in remote Oceania.

The analysis of the Papuan component shows that populations with Papuan-like ancestry have broadly similar proportions of Denisovan ancestry within their Papuan component, although their proportion of Papuan ancestry varies (Figure 4B). Exceptions include certain Island Southeast Asian populations, which show lower Denisovan ancestry proportions in their Papuan component, and the Aeta and Agta, which exhibit higher levels of Denisovan ancestry in that component. In contrast, analysis in the Austronesian component shows no detectable Denisovan introgression in most of Oceanian populations, except for a small amount $<1\%$ in Maumere and Tikopia (Figure 4C) and the Agta and Aeta populations, which have almost 2% Denisovan introgression in their Austronesian component. The fact that the Aeta and Agta show elevated Denisovan ancestry in both their Papuan and Austronesian components, explains their overall highest levels of Denisovan ancestry. By contrast, other negrito groups such as Mamanwa and Manobo show much lower global Denisovan ancestry, because the Denisovan ancestry in their Papuan component is largely diluted by their large Austronesian component, which itself has little or no Denisovan ancestry. The elevated Denisovan signal in both the Papuan and Austronesian components in the Aeta and Agta populations, a pattern not seen in the other populations, is consistent with an additional pulse of Denisovan admixture unique to these groups[53].

Together, the Denisovan signal within both the Papuan and Austronesian component is consistent with reports of multiple pulses of Denisovan introgression into modern humans in different regions and different historical ancestries[10,51,53,54]. Moreover, the near complete disappearance of these regional differences when conditioning on Papuan ancestry segments suggests that variation in Denisovan ancestry across Oceania primarily reflects differences in the amount of Papuan ancestry along the genome. This is consistent with prior reports of a strong correlation between Papuan ancestry and Denisovan introgression across Island Southeast Asia and Oceania[10,49–51,55].

### Variants of biomedical relevance in Oceania and its surrounding regions

Research on clinically relevant genetic variation in Oceania remains limited due to the under representation of populations from the Pacific in global genomic databases[56]. Although past regional initiatives have aimed to characterize biomedical variation in the region, most of the available data is sparse or limited to small cohorts[57–61]. The OGVP dataset provides a unique opportunity to explore the geographic distribution of clinically relevant variants across the Pacific and to evaluate their potential implications for precision medicine. To this purpose, we identified 47 variants of biomedical relevance in OGVP exhibiting distinct allele frequency (AF) patterns across population groups relative to the OGVP-wide mean (Figure 5). These variants were defined by integrating high-confidence ClinVar annotations[62] and pharmacogenomic loci reported by the GenomeAsia Consortium[63]. Of these, 31 variants were found to be related to the response and metabolism of different drugs, while 16 variants where found to be classified as pathogenic or likely pathogenic (SM-Tables - Variants of biomedical relevance 1–4).

For AF comparisons, neighboring populations were merged into 41 OGVP population groups (with group sample sizes ranging from 8 to 32 individuals, with a median sample size of 19; Table S3). Across these groups, we found allele frequencies to vary widely, revealing multiple highly differentiated variants (Figure 5). These differentiated variants are associated with the metabolism or response to drugs used for a broad range of clinical purposes, such as immunosuppression and inflammation, anti coagulation, infectious disease, lipid metabolism, alcohol metabolism, pulmonary disease, substance-use disorders, and cardiovascular conditions[64].

We observed pronounced between-region differentiation, for example in the simvastatin-associated variant *rs35350960*, which is substantially more common in Polynesian populations (AF = 0.23–0.40) and rare elsewhere in OGVP. This is consistent with Polynesian-specific founder effects. Similarly, *rs122777823* is enriched in Melanesian populations, with markedly higher allele frequencies than in other regions. Beyond these broad regional contrasts, we also observe differences even within regions, that is allele frequencies varying amongst neighboring island

groups—for example, Kiribati shows an elevated frequency of the methotrexate-associated variant *rs4673993* (AF = 0.72), and Chuuk shows a distinct pattern at *rs4148323*.

Overall, OGVP highlights that Oceania cannot be treated as a single reference population for clinical genetic guidelines. Deep Papuan and East Asian (Austronesian) ancestry, together with strong founder effects, generate substantial allele-frequency differences both between and within regions (Figure 5). This structure suggests the development of island and population-specific thresholds and pharmacogenomic recommendations, motivating further genomic sequencing efforts in Oceania to develop precision genomic recommendations and to guide dosing and drug-safety decisions.

## Materials and Methods

### Oceanian genotype data and general quality control

981 individuals from 92 populations and 30 countries across Oceania and Southeast Asia were genotyped on the MEGA-EX Illumina platform (1,895,184 markers before quality control) using the b37 (hg19) version of the human genome. While acknowledging that geographic classification of the Pacific might not fully represent cultural or genetic differences, for the sake of clarity we used the following labels for the major geographic groupings of our dataset: Taiwan, Mainland southeast Asia (MSEA), Island southeast Asia (ISEA), Melanesia, Micronesia and Polynesia (see Figure 1).

For quality control purposes, we separated the individuals into two groups, depending whether they have a Melanesian geographic origin ($n$=347) or not ($n$=634), to perform the following steps with PLINK 1.9[65]:

- Remove individuals with a missing call rate greater than 10% (`--mind 0.1`)
- Exclude sites with a call rate lower than 5% (`--geno 0.05`)
- Flip alleles at sites that are flipped from the definition in the human reference genome b37.
- Calculate relatedness ($r$, `--genome`) between all pairs of individuals to find and remove one member of the related pairs with $r$ greater than 0.25
- Filter SNPs with a Hardy-Weinberg equilibrium exact test p-value below 1e-8 (`--hwe 1e-8`)

Finally, we removed duplicated variants. The final working dataset (OGVP) contains 913 individuals and 1,797,365 SNPs.

### Datasets description

For the analyses that required continental references, we used the following individuals from the Human Genome Diversity Project (HGDP) panel genotyped in the same MEGA platform[66]:

- America: Surui (n=10), Karitiana (n=18), Colombian (n=9), Puno (n=20)
- Europe: French (n=28), Italian (n=12), Tuscan (n=7)
- Africa: Bantu (South Africa/Kenya) (n=19), Mandenka (n=22), San (n=60), Yoruba (n=22)
- East Asia: Cambodian (n=10), Dai (n=10), Han (including Northern Han Chinese) (n=44), Japanese (n=29), Lahu (n=8), Naxi (n=9)
- South Asia: Sindhi (n=24), Kalash (n=23)

We also built a panel of ancient DNA individuals from the geographic areas of interest that was merged with our panel of present-day individuals. Finally, we generated another dataset with the OGVP individuals and whole genome sequences from Altai Neanderthal[67] and Altai Denisova[68] to look at possible introgression from hominins in the Pacific. See Table S1 for the detailed information of the individuals used in this study.

### Patterns of genetic diversity

Since linkage disequilibrium affects Principal Component Analysis (PCA) and genetic clustering methods, we pruned the OGVP dataset with PLINK `--indep-pairwise 50 5 0.8` for these analyses. We performed a PCA with our modern and ancient individuals using the projection option of the tool `smartpca` part of AdmixTools[48]. We projected the ancient individuals onto the modern individuals and enabled the flag `lsqproject` to consider the large amounts of missing data present in the ancient individuals (Figure 2A).

Additionally, we used the program ADMIXTURE[12] to determine genetic clusters using unsupervised clustering with the number of clusters K ranging from 2 to 17. We ran 10 replicates of each of these K values to confirm stability of clustering and performed cross-validation find the best fitting K value, pong[69] was used for visualization and to identify the most representative replicate (Figures S1 and S2).

We also calculated the population pairwise Fst with `smartpca`; the results can be seen in the supplementary material (Figure S4).

### Ancestry specific analyses

In order to identify local ancestry segments in the individuals in the OGVP cohort, the haplotypes were first phased with the program SHAPEIT2[70], and then we used the software RFMix v1.5.4[71]. For this analysis, we defined four different ancestries: African-like, European-like, Papuan-like and Austronesian-like. The last two groups were made up with populations from the OGVP dataset:

- Papuan-like: Kundiawa, Tari, Mendi, Bundi, Marawaka, Enga, Western and Eastern Highlands
- Austronesian-like: Atayal, Paiwan, Bunum, Ami, Kankana-ey, Vietnamese, Han and Dai

We note that the Vietnamese and Han are not Austronesian populations, but are Austronesian-adjacent in that they are similar genetically and much more similar to Austronesian than to the other ancestries used (African, European, or Papuan)[72]. Thus, we use them to augment our reference panel in order to avoid imbalance in training samples between the four reference sets.

### Patterns of genetic relatedness

We identified identity-by-descent (IBD) segments with refinedIBD[73] on the OGVP phased data. We followed the authors' recommendations to merge short IBD segments, fill in and remove gaps between IBD segments with at most one discordant homozygote site and that are less than 0.6 cM in length. After merging, we kept only IBD segments with length $> 3$ cM and removed any segments that overlap with regions of low complexity (namely telomeres and centromeres). To generate ancestry-specific IBD calls, we intersected pairwise IBD segments with local ancestry calls inferred using RFMix. We developed a custom pipeline that assigns a local ancestry label to each IBD segment by matching the genomic coordinates of the IBD tract with the corresponding local ancestry calls in both individuals. We used only Papuan and Austronesian IBD segments for all downstream IBD analyses by summing the length of shared Papuan and Austronesian IBD segments over all pairs of individuals in each population and normalizing by the Papuan and Austronesian global ancestry fraction.

To further investigate the genetic relationships of the Polynesian Outliers, we utilized the ancestry-specific f3. Ancestry-specific allele frequencies were estimated by masking genotypes according to local ancestry assignments inferred with RFMix. For each ancestry, only haplotypes assigned to that ancestry were retained, while all others were masked as missing. Allele frequencies were then calculated independently for each ancestry using the masked genotype data. These ancestry-specific allele frequencies were subsequently used to compute ancestry specific f3 for Papuan and Austronesian haplotypes using `f3` in snputils[74]. This statistic is used to measure the differences in allele frequency in test populations (A and B) in reference to an outgroup (C): f3(A, B;C). In this case, we considered the Polynesian Outliers (Kapingamarangi, Ontong Java, Rennell & Bellona, and Tikopia) as test populations B against all OGVP populations as test populations A and the Andaman Islands as the outgroup C.

### Comparison with ancient DNA

In the case of the ancient individuals from the Pacific, we focused on eight individuals identified as part of the Lapita culture in Tonga and Vanuatu (see Table S2). We computed the allele frequencies for all the SNPs present in at least one of the ancient individuals (141,425 sites) and separated the individuals according to their geographic origin: Lapita-Vanuatu (MAL006,I5951,I1369,I1368 and I1370)[23–25] and Lapita-Tonga (TON001, TON002 and CP30)[23,24].

We estimated the population divergence using an ancestry specific approach, where we calculated the average pairwise differences ($\pi$) between the ancient Lapita individuals and populations from the OGVP cohort separately in the Papuan and the Austronesian components of the latter. Specifically, we used previously defined Papuan or Austronesian segments from RFMix and calculated all the possible pairwise differences between the populations to generate an ancestry specific neighbor joining tree using the `bionj` function part of the `ape` R package[75] (Figures 2C and S9).

### Comparison with archaic DNA

We used the program `snpflip` to flag any reversed or ambiguous sites in the HGDP, OGVP, and archaic files (both Neanderthal and Denisova). We flipped all the reversed sites and removed all of the ambiguous SNPs using PLINK (`--flip` and `--exclude`). We then kept only-biallelic sites in the human datasets (OGVP and HGDP), and merged them on those sites. Finally, we extracted the sites in this merge from the archaic hominin data, removed triallelic sites, and created a final merged file. The result was a dataset containing the information of OGVP, HGDP, Neanderthal, and Denisova.

Given the known presence of Denisova introgression in New Guinea[51], we were interested in examining if this presence spread further in Oceania. For this purpose, we applied the f4-ratio using FrAnTK[76] to quantify the hominin introgression in our dataset. The f4-ratio allows the estimation of the mixing contributions of an admixture event[48]. We used the test f4(Altai Neanderthal, Yoruba; OGVP, Han)/f4(Altai Neanderthal, Yoruba; Denisova, Han), designed to isolate Denisovan signals in populations that also have Neanderthal introgression[49] (Figure 4)A).

We then quantified the proportion of Denisovan ancestry in ancestry-specific data determined by our RFMix results. From the reference dataset, we extracted the populations to be used as references for this analysis (Yoruba, Han, Dai, Neanderthal, and Denisova). Then, we obtained the sites available for both the ancestry-specific Oceanian data and reference data and extracted those sites from both datasets. Once we were working with the same sites in both datasets, the reference dataset, which was in PLINK format, was used to build two FrAnTK[76] datasets: one containing a merge of the Austronesian ancestry-specific OGVP data and the references and a second merge of the Papuan ancestry-specific OGVP data and the references. An f4-ratio analysis was then applied to these datasets to obtain ancestry-specific estimations of Denisova introgression in the populations (Figure 4B,C).

### Variants of biomedical relevance

Variants of biomedical relevance in the OGVP dataset were identified by integrating data from the ClinVar database[62] and the GenomeAsia consortium[63]. ClinVar data (release version 1.6.62, accessed on August 16 2024) were obtained from the ClinVar FTP repository and filtered to retain variants present in OGVP with a clinical significance rating of at least three out of four stars, corresponding to variants reviewed by an expert panel. Variants reported by the GenomeAsia consortium as pharmacogenomic loci were also included if present in OGVP[63].

For each retained variant, allele frequencies (AF) of the alternative allele were calculated across 41 population groups defined within OGVP based on admixture profiles and geographic proximity using PLINK 1.9. To identify population-specific frequency differences, we compared each group's AF to the weighted mean AF across all OGVP groups, where the weights correspond to the number of observed chromosomes ($N_{\text{CHROBS}}$) per group, accounting for unequal sample sizes across population groups. A two-proportion Z-score was then computed for each population group–variant combination.

Variants with absolute Z-scores greater than 3.29 (corresponding to a two-sided $p < 0.001$) were considered significantly differentiated. To provide broader context, AFs for the same variants were retrieved from three continental population cohorts in gnomAD v2.1.1 (East Asian, African/African American, and non-Finnish European), which were included as external references but excluded from the significance test. Only variants with AF greater than 0.1 in at least one OGVP population were retained for visualization (Figure 5, Table S3).

## Acknowledgements

We thank the participants who contributed samples from the populations included in this study. We recognize the leadership of John Clegg in assembling the collection of materials that enabled the foundation of the Oceanian Genome Variation Project (OGVP). We appreciate the assistance of Michael Alpers, former head of the Papua New Guinea Institute of Medical Research and Professor Sir David John Weatherall (deceased), founder of the Medical Research Council Weatherall Institute of Molecular Biology at the University of Oxford. We thank Euan Ashley and the Oxford-Stanford Big Data Initiative for supporting data generation in partnership with Illumina Inc. We thank Evelyne Heyer from the Musée de l'Homme in Paris, France, for providing sampling materials and making a subset of samples from Luzon available for this work. We thank the Vanuatu Kaljoral Senta/Cultural Centre and Director Richard Shing. We also thank Kalina Kassadjikova, Sarah Kaewert and Chris Gignoux for collaborative input.

## Data availability

Data will be made available to researchers for population genetics studies through the European Genome-phenome Archive (EGA) upon publication.

## Ethics statement

Individual samples analysed in this study were collected over a period of approximately three decades in close collaboration with local investigators, institutions, and community partners. At the time of collection participants provided informed consent for genetic research, long-term storage of biological material, and use in population genetics studies, in accordance with the contemporary ethical standards and regulatory requirements. All samples were anonymised prior to analysis. Approvals were obtained from relevant national and institutional review bodies, including the appropriate health authorities in the countries of origin and have received updated reapprovals from the University of Oxford Tropical Research Ethics Committee (OxTREC; approval references and the University of Oxford Tropical Research Ethics Committee (OxTREC; approval references 537-14 and 566-24).

We recognise that ethical norms and expectations regarding genomic research, data sharing, and the stewardship of biological materials have evolved substantially over the past three decades. Where feasible, we have undertaken extensive efforts to re-engage with originating investigators, institutions, and relevant community representatives to seek renewed agreement for contemporary genome-wide analyses, and to inform, guide and review the proposed analyses. These consultations have included discussion of study aims, data generation, governance, and plans for dissemination. We acknowledge that recontact is not always possible for legacy collections and that historical consent frameworks may not have fully anticipated all current genomic science consent frameworks.

The populations represented in this study have historically been under-represented in global genetic research. Our intention in generating these data is to improve equitable representation and scientific understanding while adhering to principles of respect, transparency, and partnership. We are committed to ongoing dialogue with regional stakeholders and to supporting locally led governance structures for genomic data and sample stewardship, consistent with emerging frameworks for Indigenous and Pacific data sovereignty. Where appropriate and feasible, we have committed to work with partner institutions to facilitate local capacity building and the future custodianship or shared governance of biological materials and derived data.

We acknowledge that the ethical landscape surrounding the use of legacy biological samples remains complex and evolving. We remain open to continued consultation and review to ensure that the conduct and dissemination of this research align with the values and expectations of the communities from whom these samples originate.

## Author contributions

Conceptualization: A.G.I., A.M.E., and A.J.M. Ethics approval: A.J.M., T.P., K.A., K.R., and A.V.S.H. Formal analysis: C.D.Q.C., C.B.J., P.A.G., S.V.S., P.R.R., R.G.B., J.R.H., J.V.M.M., A.L.S., G.L.W., and J.B.P. Interpretation of results: A.G.I., A.M.E., S.J.O., C.D.Q.C., C.B.J., P.A.G., S.V.S., P.R.R., R.G.B., J.R.H., A.L.S., F.D., and M.C.A.D.U. Writing and editing: P.A.G., C.D.Q.C., C.B.J., P.R.R., R.G.B., and A.G.I., with approval from all co-authors. Supervision: A.G.I., A.M.E., and A.J.M. Funding acquisition: A.G.I., A.M.E., A.J.M., and C.D.B. Sample collection and preparation: K.A., K.S., K.R., I.V., A.S.M.S., G.K., G.A., J.S.F., F.F., A.A., S.A., A.V.S.H., T.K., P.E., M.Sp., M.St., W.P., M.P., and S.J.O.

## Funding

Data generation was supported by the Oxford-Stanford Big Data Initiative. C.D.Q.C. was supported by a CONACYT Postdoctoral Fellowship and a CINVESTAV visiting professor scholarship. C.B.J., S.V.S., and R.G.B. were supported by CONACYT graduate scholarships. P.A.G. was supported by Stanford Graduate Fellowship and Fulbright. C.D.B. would like to acknowledge support from the Chan Zuckerberg Institute (CZI). P.E. was supported by the European Union's Horizon 2020 research and innovation programme under the Marie Skłodowska-Curie grant agreement No 873207.

## Declaration of interests

The authors declare no competing interests.

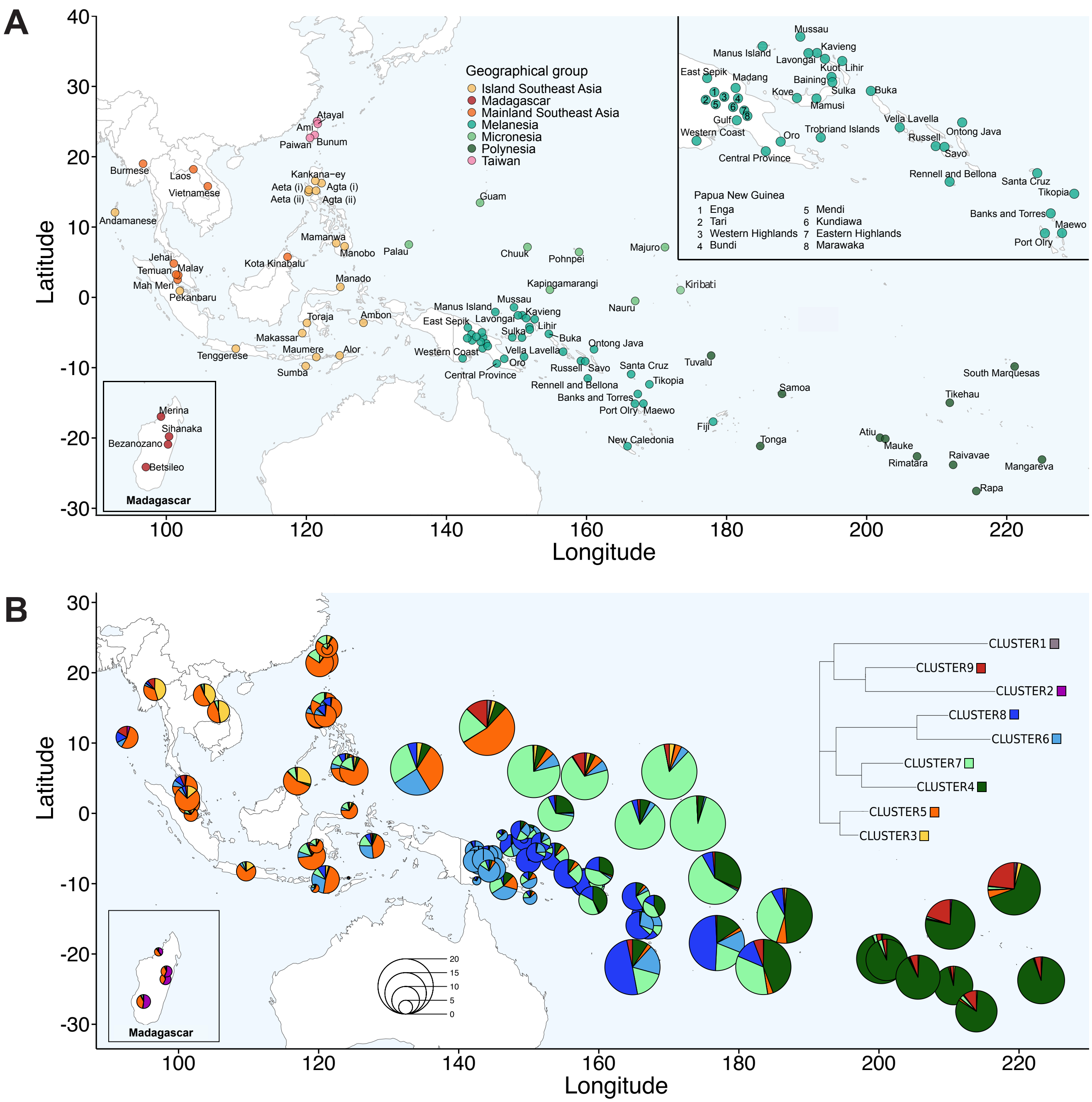

**Figure 1. A.** Geographic distribution of all the individuals used in this study. Each circle represents a population colored by its geographic location. Papuan and neighboring populations are shown in an inset upper right for clarity. **B.** Pie charts show the average ancestry proportions for each location at K=9 using ADMIXTURE unsupervised clustering with the size of each pie-chart proportional to the population's sample size. The hierarchical clustering tree of the nine unsupervised clusters, based on the inferred allele frequencies of each cluster, is shown at upper right.

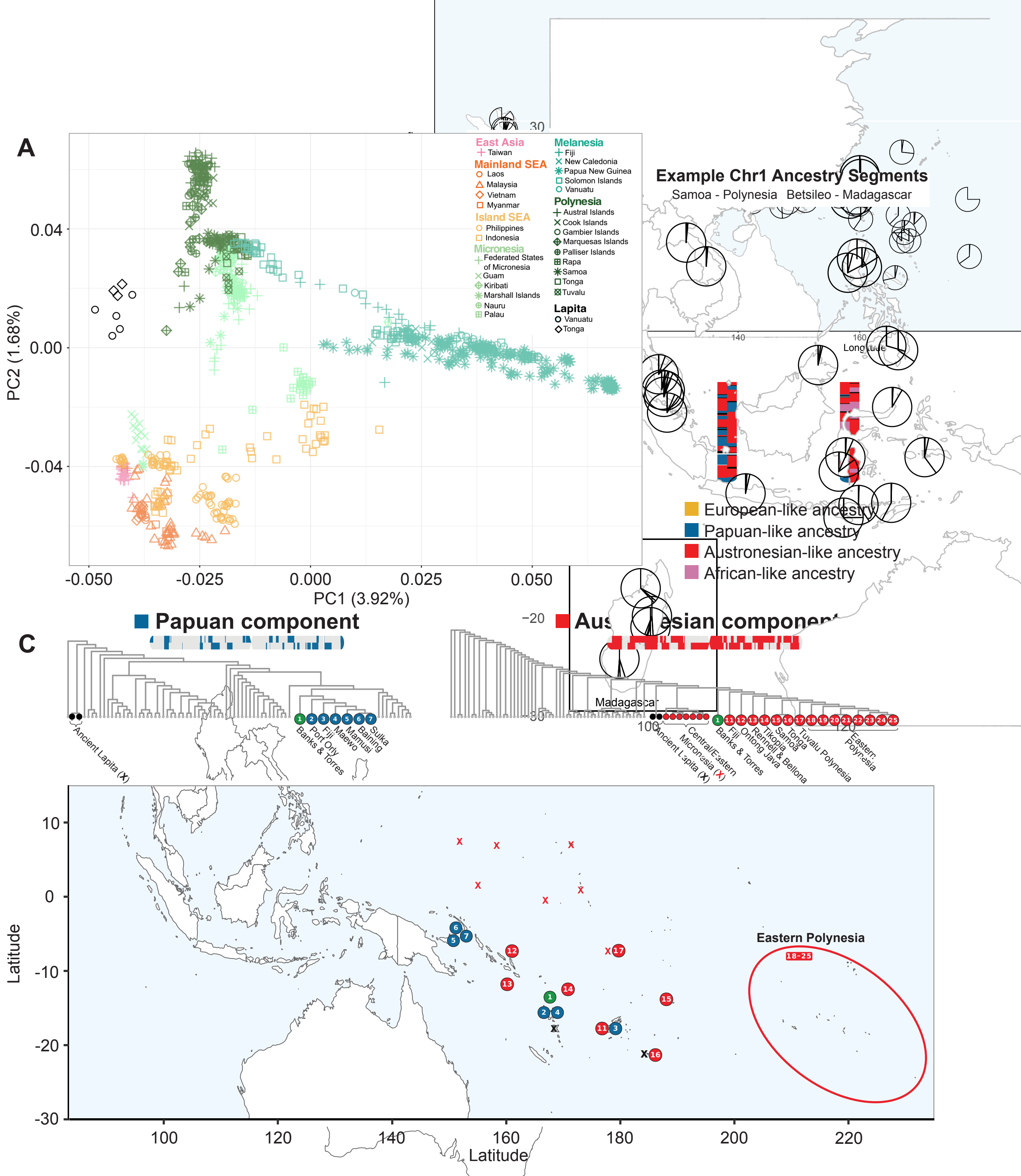

**Figure 2.** Ancient Lapita derived ancestry in present day individuals. **A.** Principal component analysis of the OGVP dataset and ancient individuals from the Pacific. Ancient Lapita samples are represented as black symbols. **B.** Local ancestry (segment-level ancestry) of two individuals from the OGVP cohort. **C.** Neighbor-joining tree of the Lapita sequences from Tonga and Vanuatu together with the Austronesian ancestry-specific chromosomal segments from the modern individuals. The map shows the geographical location of all highlighed populations from the neighbor-joining tree.

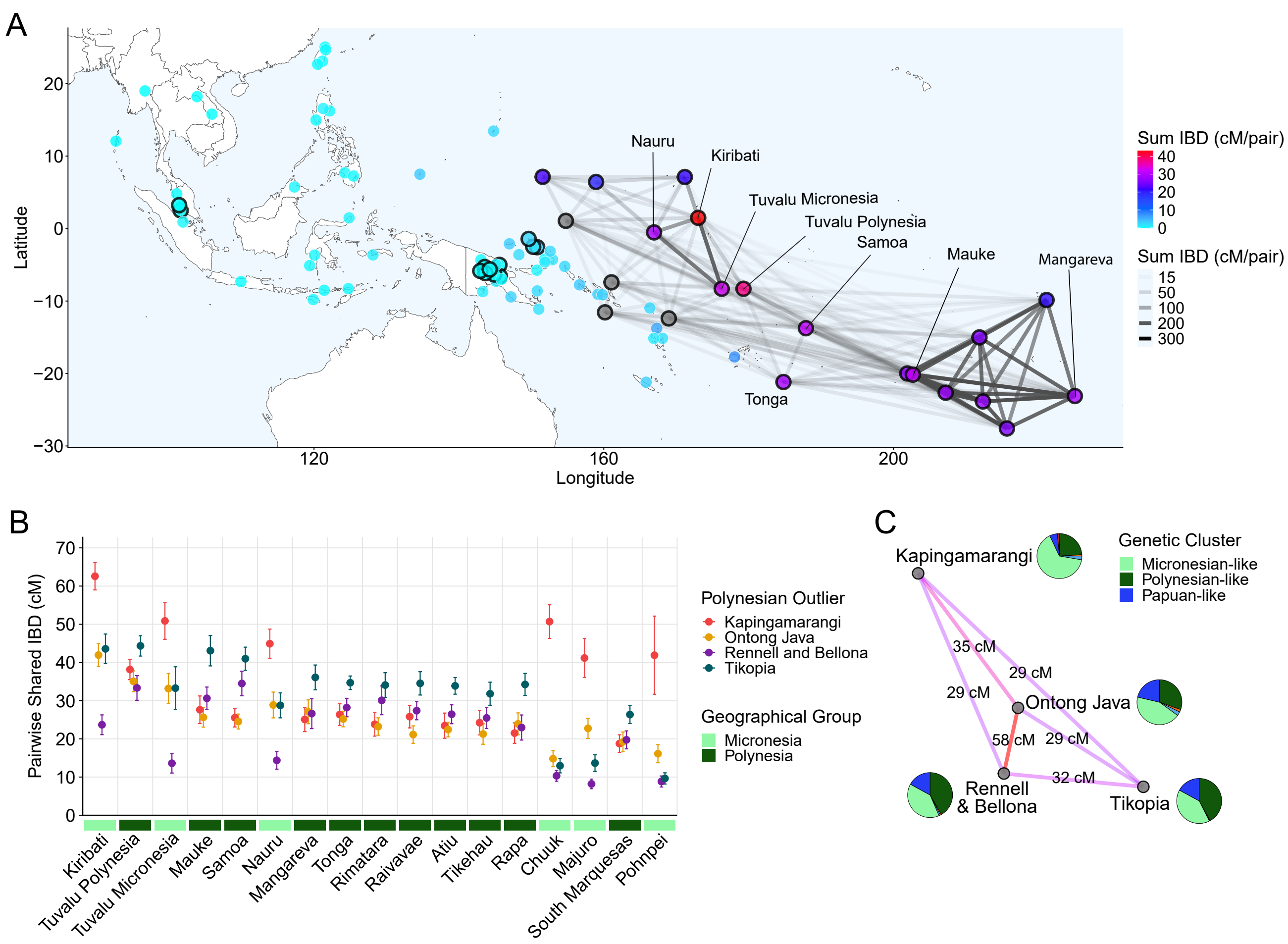

**Figure 3.** **A.**, Network based on the amount of IBD sharing between populations. The amount of shared IBD is shown as an edge between locations and the color of each edge is relative to the average cM of shared IBD. Only edges with values > 10 cM are shown, and nodes connected to at least one other node are outlined in black. The nodes of the network are colored according to the mean value of the shared IBD between the Polynesian Outliers and the population. **B.**, Scatterplot showing mean shared IBD between each of the four Polynesian outliers and Polynesian and Micronesian populations. **C.**, Network based on the amount of IBD sharing between the Polynesian Outliers. Edges are colored similarly to A. Pie charts show ancestry proportions for the Polynesian Outliers from Figure 1B.

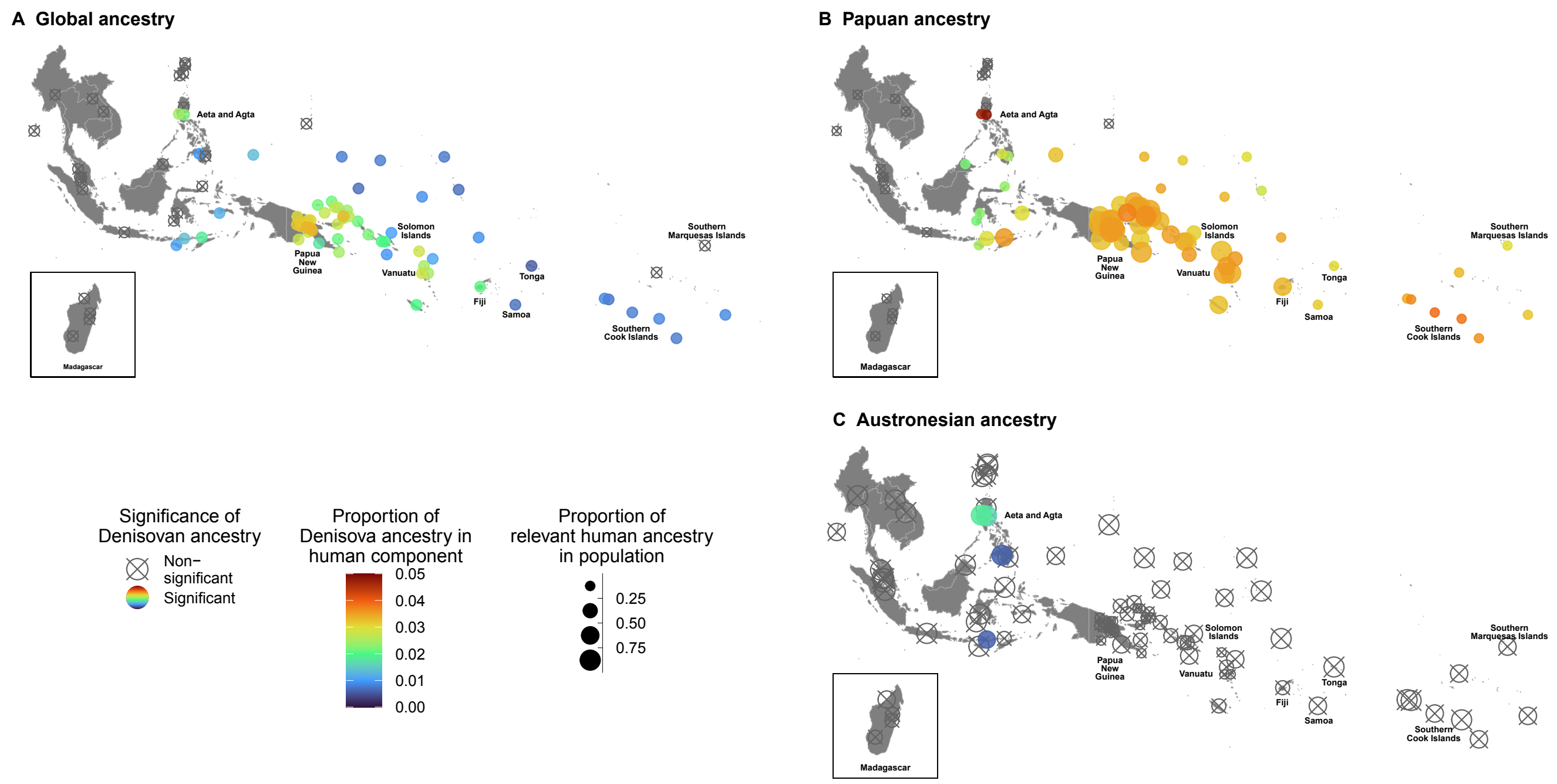

**Figure 4.** Proportion of Denisovan ancestry estimated by f4 ratio across the entire genome (**A**), in Papuan ancestry segments (**B**), and in Austronesian-like ancestry segments (**C**). Symbol shape distinguishes significant from non-significant Denisovan signal. The symbol color in populations with significant signals indicates the proportion of Denisovan signals in the ancestry component. Symbol size in panels **B** and **C** represents the average proportion of the human ancestry component in each population.

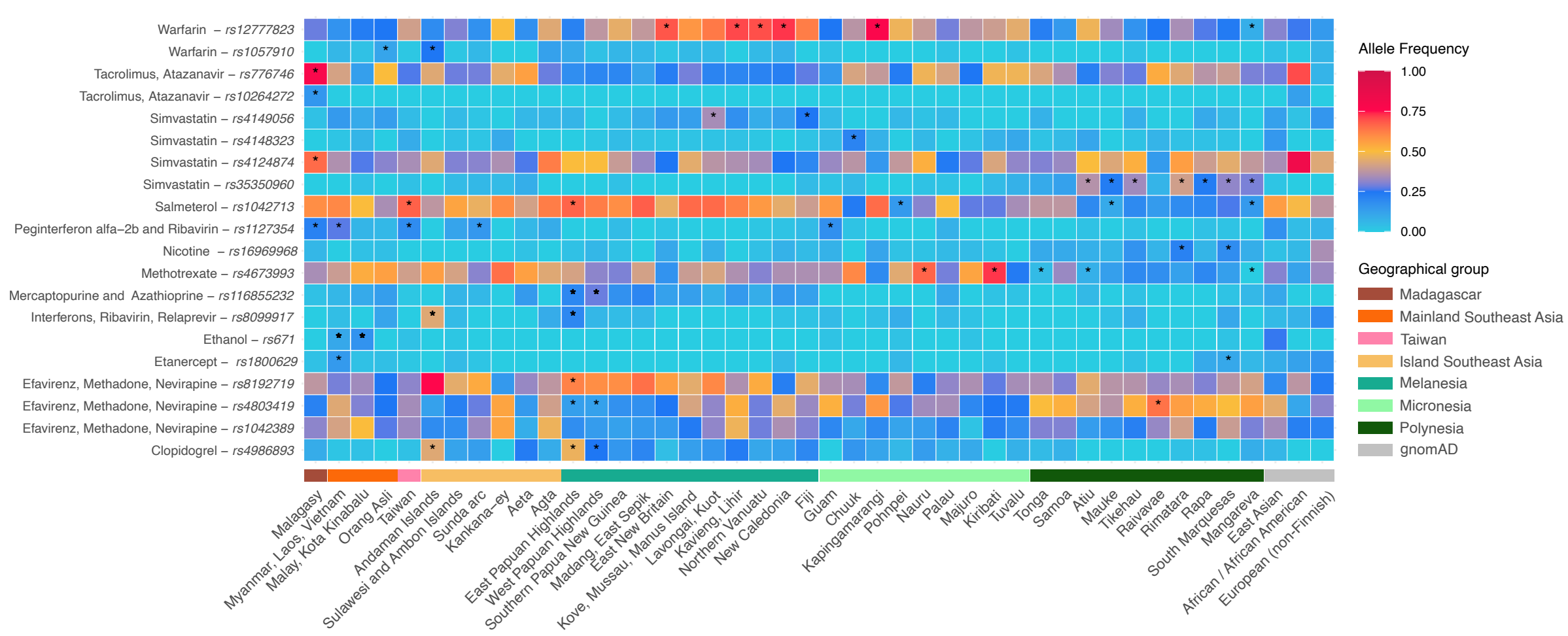

**Figure 5.** Allele frequency heatmap of variants involved in drug response. Each tile represents the alternate allele frequency (AF) of a given SNP (y-axis) across population groups in OGVP (x-axis), with continental reference populations from gnomAD. Tile colors range from blue (low AF) to red (high AF). Asterisks (*) denote population–SNP combinations that deviate significantly ($p < 0.001$) from the weighted mean AF across OGVP populations. Only variants with an AF greater than 0.1 in at least one OGVP population are shown. Clinical annotations were obtained from ClinVar[62], and from pharmacogenomic loci reported by the GenomeAsia Consortium[63].

# Supplementary material for The Genomic Landscape of Oceania

**Consuelo D. Quinto-Cortés*[1], Carmina Barberena-Jonas*[1], Peter A. Gerlach*[2], Sofía Vieyra-Sánchez[1], Ram González-Buenfil[1], Petria R. Russell[3], Stephen J. Oppenheimer[4], Kathryn Auckland[5], Kathryn Robson[4,27], Tom Parks[4], Julian R. Homburger[2], Alissa L. Severson[2], Genevieve L. Wojcik[6], J. Víctor Moreno-Mayar[7], Javier Blanco-Portillo[8], Karla Sandoval[1], Ishwar Verma[9,27], Abdul Salam M. Sofro[9], Frederick Delfin[10], Maria Corazon A. De Ungria[10], George Koki[11], George Aho[12], Jonathan S. Friedlaender[13], Francoise Friedlaender[13,27], Angela Allen[14,15], Stephen Allen[15,16], Adrian V. S. Hill[4], Toomas Kivisild[17], Phillip Endicott[18,19,20,21], Matthew Spriggs[22,23], Mark Stoneking[24,25], Carlos D. Bustamante[2], Maude Phipps[26], William Pomat[11,**], Alexander J. Mentzer[4,**], Andrés Moreno-Estrada[1,**], and Alexander G. Ioannidis[2,3,**]**

*These authors contributed equally to this work
**Correspondence: william.pomat@pngimr.org.pg, alexander.mentzer@ndm.ox.ac.uk, andres.moreno@cinvestav.mx, ioannidis@stanford.edu

## Supplementary material

## Supplementary Figures

## Supplementary Tables

# 1 Unsupervised clustering

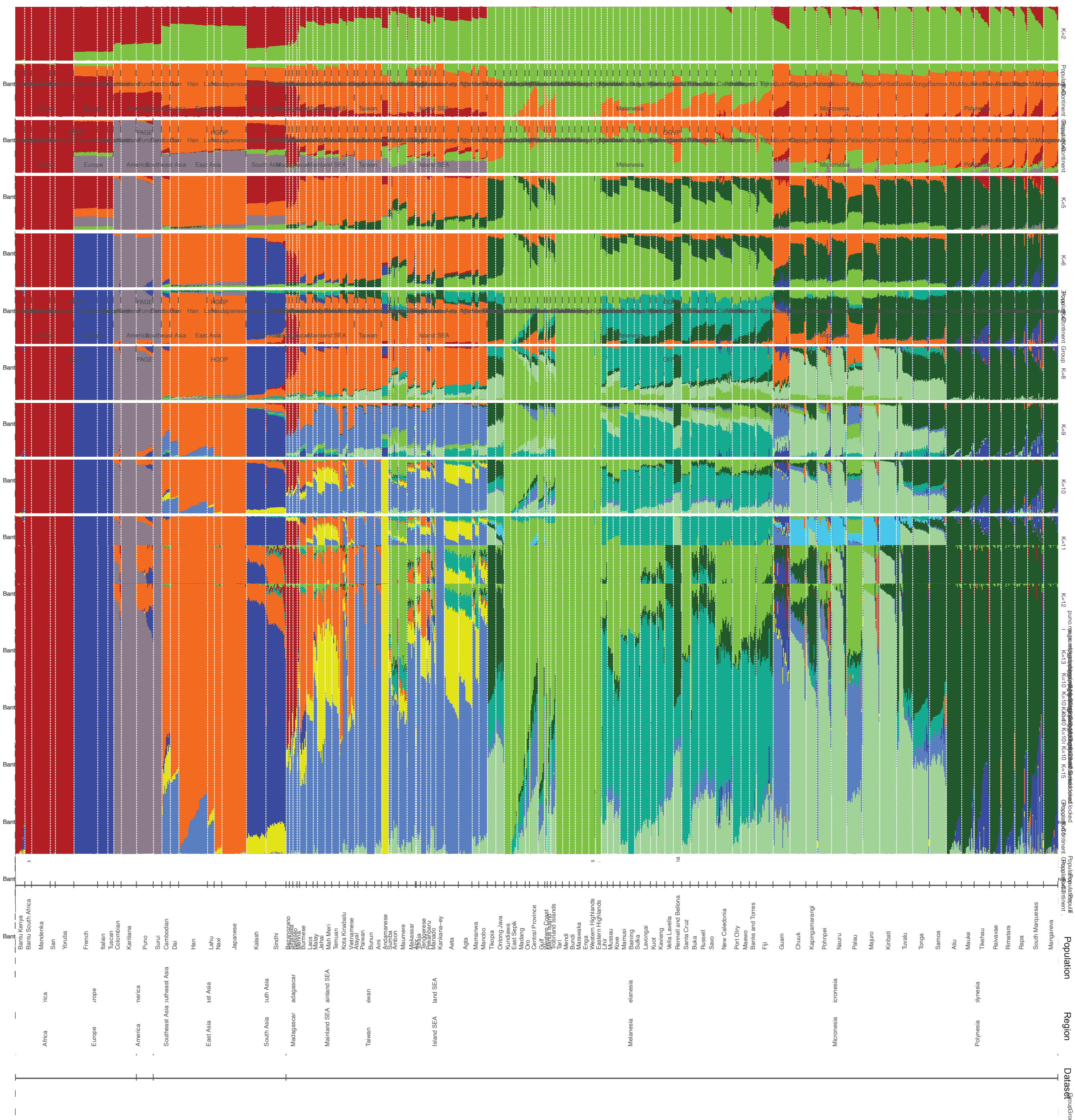

**Supplementary Figure S1.** ADMIXTURE plots from K=2 to K=17. Genetic clustering was performed with ADMIXTURE[1], and we show the dominant mode for each K as identified by pong[2].

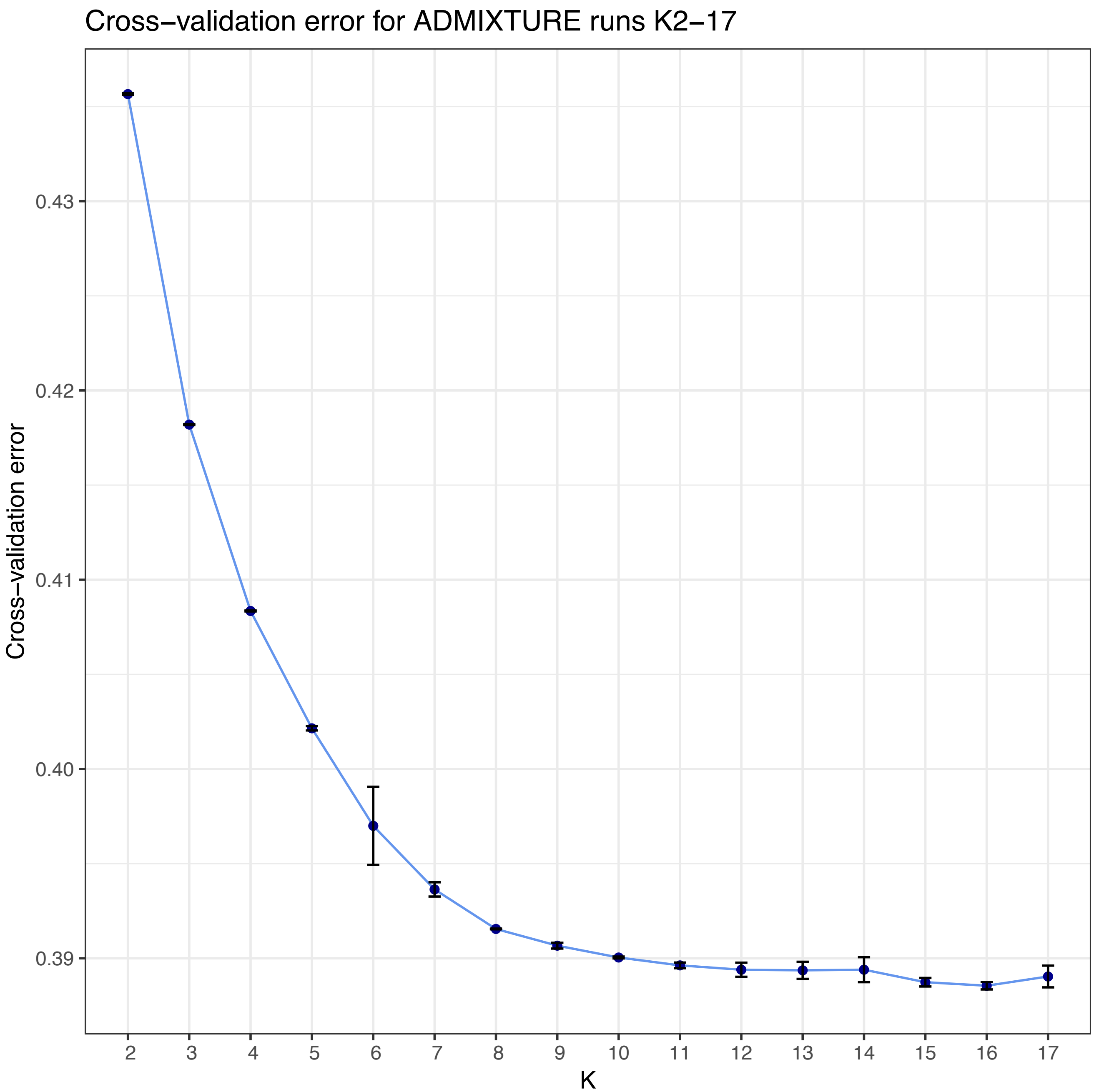

**Supplementary Figure S2.** Cross-validation with ten replicates of ADMIXTURE for each K. The point represents the replicates' mean and the bars the standard deviation.

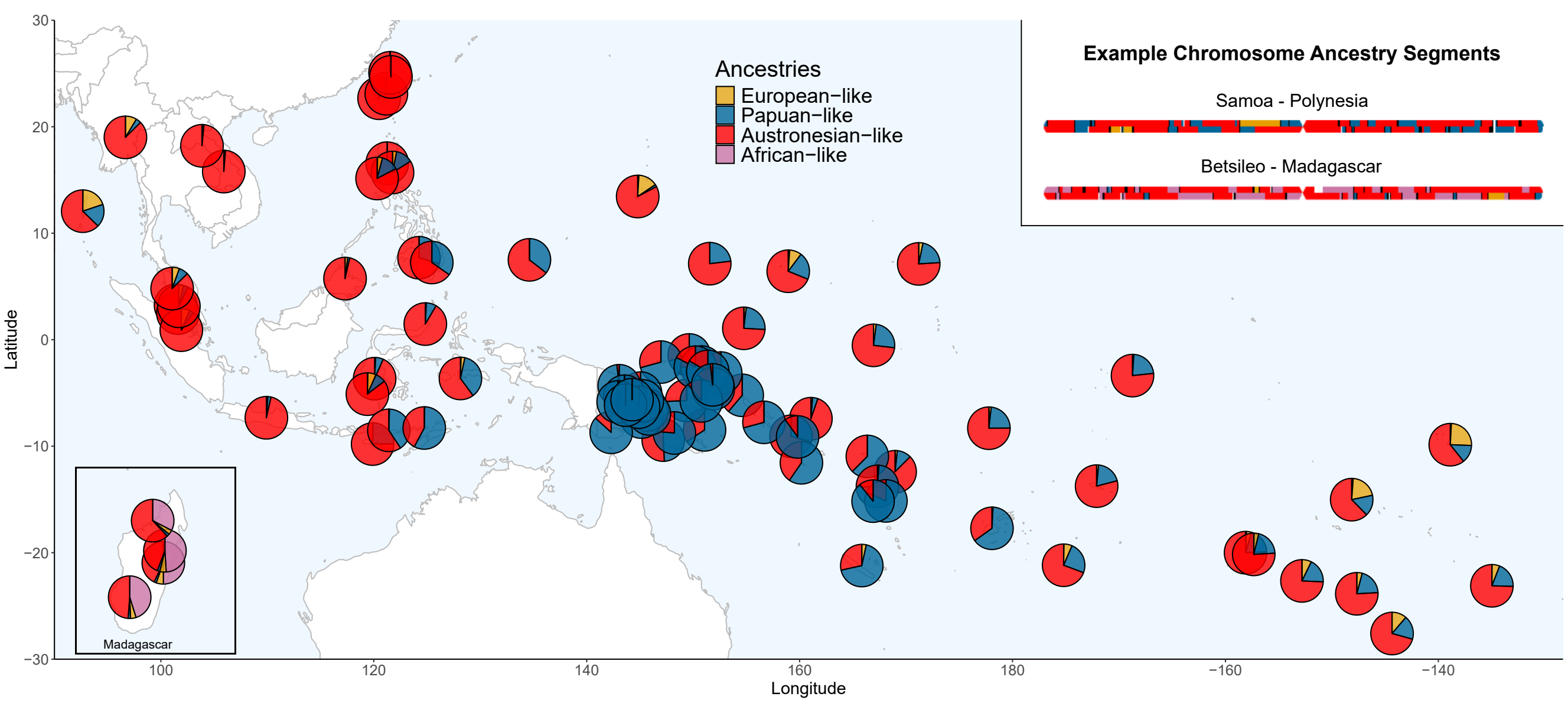

**Supplementary Figure S3.** Proportion of windows from European-, Papuan-, Austronesian-, and African-like ancestry aggregated across all individuals in each population. Window-level ancestry was inferred by RFMix, and illustration of those local ancestry calls in an example individual is shown in the upper right panel, which reproduces Figure 2B.

## 2 Notable regional patterns in ADMIXTURE clustering

In addition to the broad regional structure described in the main text, the region-wide ADMIXTURE analysis reveals several notable region-specific patterns. Here, we highlight patterns in Palau, Guam, and along the southern coast of New Guinea that show further complexity to the broad regional trends.

### 2.1 Palau and Guam

Palau and Guam both harbor several ancestry components at high proportions in the region-wide ADMIXTURE analysis in contrast to the other Micronesian populations (Figure 1). In Palau, we observe substantial East Asian–like, Papuan–like, and Micronesian-like ancestry. Recent ancient DNA work shows that the earliest sampled Palauans already carried an ancestry profile similar to that of present-day Palauans, with approximately 60% East Asian-related and 40% Papuan-related ancestry[3]. The presence of Micronesian-like ancestry in Palauan individuals in the OGVP dataset points to additional affinity with populations contributing to this broader regional component. Consistent with this, Palau individuals fall in an intermediate position between Papuan, Southeast Asian, and Micronesian/Polynesian clusters in a PCA (Figure 2). For Guam, we observe predominantly East Asian–like ancestry, with additional European- and Native American–like components, in agreement with previous work and likely reflecting admixture associated with the Spanish colonial Manila galleon trade linking Guam to both the Philippines and the Americas[4,5]. As in Palau, we also observe Micronesian-like ancestry in Guam.

### 2.2 Southern coast of New Guinea

The coastal populations on the southern coast of New Guinea show another case of populations with distinct ancestry components compared to neighboring populations. In addition to the expected Papuan-like ancestry, these populations show appreciable East Asian–, Micronesian-, and Polynesian-like components. These Austronesian-speaking groups have previously been described, at low values of K (for example, K = 2), as a mixture of deeply diverged Papuan and Southeast Asian ancestries[6]. Our identification of Micronesian- and Polynesian-like components refines this picture and suggests additional gene flow from populations related to present-day Micronesians and Polynesians, potentially via back migrations or long-distance exchange networks along the southern New Guinea coast. Consistent with this interpretation, a recent study of ancient DNA reports the B4a1a1a mitochondrial lineage, known as the "Polynesian motif", in individuals from coastal sites on the southern coast of New Guinea[7].

## 3 Population statistics

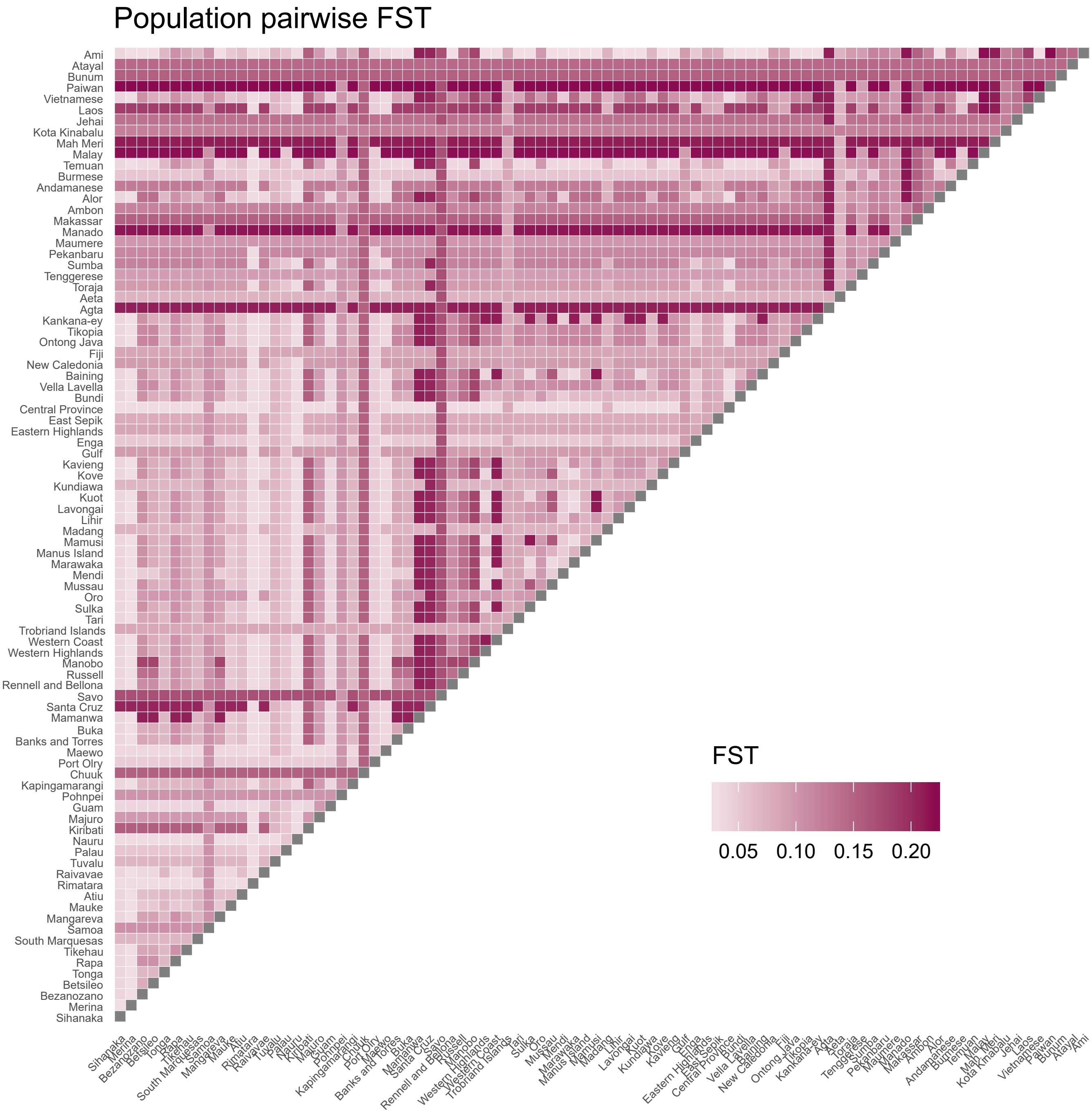

**Supplementary Figure S4.** Population pairwise Fst as calculated by `smartpca`. Darker colors represent higher Fst values.

## 4 Distribution of uniparental lineages

We used the program Haplogrep2[8] to examine the diversity of mitochondrial haplogroups, and we utilized Snappy[9] to look at the variety of chromosome Y lineages. We use the nomenclature of the International Society of Genetic Genealogy, Y-DNA Haplogroup Tree 2019-2020, Version: 15.73.

### 4.1 Mitochondrial haplogroups

The most common haplogroup seen in the Pacific is haplogroup B4, in particular haplogroup B4a1 and its derivatives (Figure S5). Among them, we find the so-called "Polynesian motif" that defines a mitochondrial lineage that is exclusively present in Austronesian-speaking populations and almost reaches fixation in Polynesia. This motif originated more than 6000 years ago near the Bismarck Archipelago. Its immediate ancestor is practically restricted to Near Oceania, and has been seen in previous studies[10].

We also observe the haplogroup Q in the Highlands and other populations of Papua New Guinea. Haplogroup P is also very frequent in the Eastern Highlands and in the Agta population from the Philippines. We observe a few occurrences of the maternal haplogroup M7c (that derives ultimately out of Taiwan around 3,000 to 4,000 years ago[11]) in individuals from Island Southeast Asia, in Pohnpei, Tuvalu, Chuuk, Kiribati, Nauru, Tikopia, Ontong Java and in one Bezanozano individual from Madagascar.

Haplogroup E is found in Guam, Madagascar, the Bismarck Archipelago, Ontong Java and Taiwan. The presence of this haplogroup in Oceania might be explained by migrations from Mainland South East Asia via Island South East Asia[11,12]. Other haplogroups found in this dataset are H, F, R, L, and Y, K, G, U, C, I, N in small percentages.

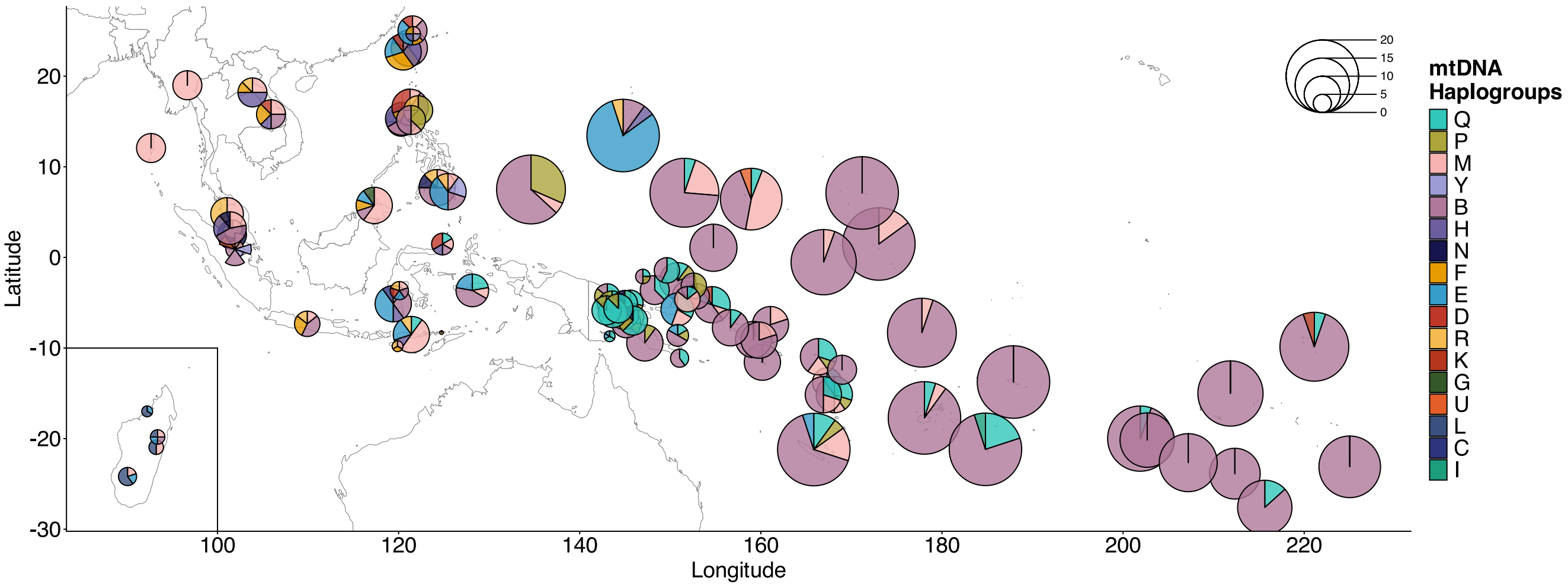

**Supplementary Figure S5.** Distribution of the maternal haplogroups in all of the populations in this study.

### 4.2 Y-chromosome haplogroups

In contrast to maternal haplogroups, we find a greater variety of paternal haplogroups and different patterns in Melanesia (haplogroups S and M), Micronesia (K and O) and Polynesia (C, R and Q) (Figure S6). Consistent with the ADMIXTURE results in Polynesia, the European and Native American influences are evident from the presence of R and Q haplogroups.

Haplogroup C is thought to be one of the original haplogroups present in Oceania, explaining its ubiquity in the region[13]. Consistent with previous findings, we observe C1a2a (V182) localized to Southeast Asia and derived subclades C1b2a (M38) and C1b2a1 (M208) are restricted to Papuan New Guinea, Melanesia, Micronesia, and Polynesia[14,15]. It has been proposed that this divide suggests the derived haplogroups originated in Melanesia[14]. In Polynesia, C1b2a1 is by far the most abundant haplogroup on all islands. This suggests that either the settling populations were largely C1b2a1, or this haplogroup has replaced the haplogroup of the settling populations.

Like C, haplogroup K is also thought to be one of the founding haplogroups in Oceania, which leads to its prevalence throughout the region[14]. Our findings are consistent with previous works[14–16]. Specifically, K2b1 (P397/399) exists at low frequencies in Indonesia and Malaysia, high frequencies amongst Filipino native groups, and throughout Oceania, with high frequencies in Micronesia[14,16].

Haplogroup S is thought to have originated in New Guinea and spread outward[14,15]. In accordance with this, it is most common and concentrated in New Guinea and in some individuals from Micronesia. We observe haplogroup S1a1b (M320) in Melanesia as well as Indonesia and in one Y chromosome in Guam. Lineage S1a2a (P307) is found across Melanesia and Micronesia.

Haplogroup M is a subgroup of K, and has previously been observed at low to moderate frequencies in Indonesia and Malaysia, and high frequencies in Papua New Guinea and Melanesia[14–16]. M (SK1828/S322) is most concentrated in Papua New Guinea. M1 (M5) is found in one individual from Alor, one individual from Tonga and in a number of Highlanders from Papua New Guinea. M1a1a2a (P87) is found in coastal Papua New Guinea. M2 (M353) is restricted to Melanesia.

Haplogroup P (P295) has previously been reported in Aeta populations in the Philippines[16]. Consistent with this, we find it in those populations as well as in two individuals from Manado.

Haplogroup O probably originated in East Asia and later migrated into the South Pacific. The lineage expanded into Taiwan (high frequency in the aboriginal Taiwanese), Indonesia, Melanesia, Micronesia, and Polynesia. We see a high diversity of lineages of this haplogroup in our dataset. Haplogroups O1a1 (M119) and O1a1a (M307.1) are common in the Philippines, Melanesia and Micronesia. In Taiwan, these lineages achieve high frequency in Taiwan, and are thought to mark the Austronesian expansion and spread of agricultural technology[14,15]. Derivatives of O1b1 are found in Mainland Southeast Asia, Melanesia and Madagascar. In particular, haplogroup O1a2 (M110) has been shown to link Oceania to Taiwan[15].

Haplogroup O2 (F36) is thought to have originated in Southeast Asia and later spread through Oceania, with O2a2 (P201) marking the expansion of agriculture[14]. Derivatives of O2a2b can be found in Mainland and Island Southeast Asia, Micronesia and Polynesia.

Throughout Indonesia and Malaysia we find the presence of haplogroups common in the Middle East and South Asia, which may be due to the history of trade with Muslim merchants, and prior to that trade with Hindu and Buddhist merchants from ancient India. Two R2a haplogroups are present in Malaysia and Indonesia. Three J2a1(L27) haplogroups are present in Makassar and Jehai Negrito, however J2 may also have a European origin. A haplogroup L1a1 (M27) is present in Makassar, Indonesia and L1a2 (M357) in the Bezanozano people in Madagascar. We also find haplogroups more commonly found in Asia. We observe one N (M231) in Atiu. One F (M89/PF2746) is present in Palau. All 4 males in the Andaman Islands are D1a (CTS11577), which matches previous findings[17]. D1a2a1c1a~D-CTS1670/Z1546 is found in Chuuk (Micronesia), which is notable because D1a2a (M64.1/M55) is a Japanese group, so this may be a consequence of Japanese occupation during World War II.

Haplogroups of European ancestry (I1a, I2, R1, G2) are present throughout Oceania, Micronesia, and Indonesia, reflecting recent exploration and colonization. The R1a and R1b haplogroups are the most common and widespread, followed by I and G2b1 (M377) in Tikehau. As expected, we found 6 African Y chromosomes, almost entirely in populations from Madagascar: One B2a1a1a1 (M109) is present in the Bezanozano and E1b1a1a1a2a1 (M4254)

Y chromosomes in the Sihanaka, Betsileo, and Merina populations. One E1b1a1a1a2a1 (M4254) haplogroup is present in Guam, which is likely the consequence of recent admixture on an island with continuing admixture from the United States of which it is a territory. Finally, we find one Q1b1a1a (M3) in Rapa Iti, which originates in the Americas.

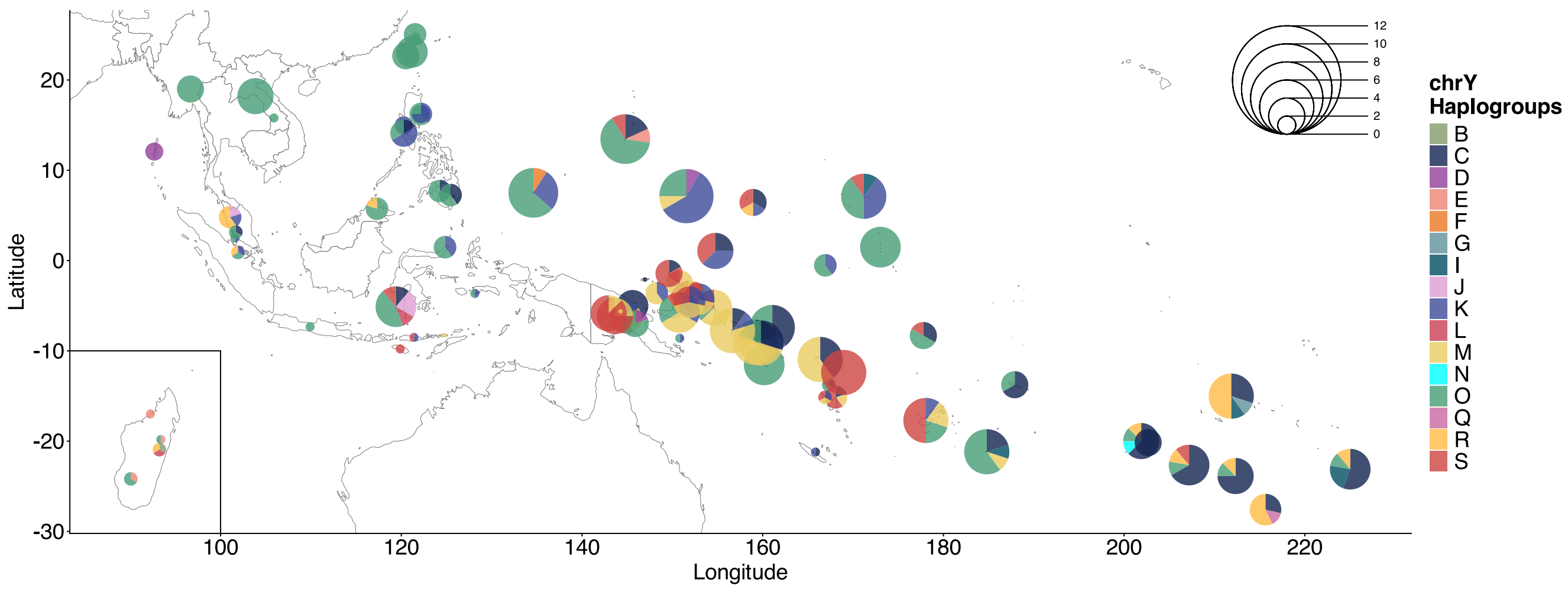

**Supplementary Figure S6.** Distribution of all the paternal haplogroups in the study.

## 5 Tuvaluan individuals' ancestry proportions

We observed that Tuvaluan individuals have ancestry proportions that are similar to those of Samoans and Tongans, with comparable contributions from Polynesian, Micronesian, and Melanesian associated ancestry components (Figure S1 and Figure S7). This suggests a shared history of Tuvalu with Samoa and Tonga, consistent with a possible settlement of Tuvalu from Samoa and Tonga[18]. However, five additional Tuvaluan individuals have predominantly Micronesian-like ancestry. This heterogeneity may reflect structure across the islands of the nation of Tuvalu. In particular, while a Polynesian language is spoken on most islands in Tuvalu, a Micronesian language is spoken on the island of Nui, which was settled by individuals from Kiribati before European contact. Consistent with this, we find that Tuvalu Micronesia-like individuals cluster most closely with Kiribati of all populations in the OGVP dataset (Figure S9) and high IBD sharing between the five Micronesian-like individuals from Tuvalu and Kiribati (Figure 3A). The population structure in Tuvalu with five Tuvaluan Kiribati-like individuals having the same admixture proportions between themselves and the remainder of the Tuvaluan individuals having different (but between themselves identical) admixture fractions, is not consistent with recent admixture, which would be expected to produce high variance in admixture fractions between individuals[19]. This suggests that these two groups are distinct and reproductively isolated and derive from older events (pre-European migration from Kiribati to Nui Island in Tuvalu) before the colonial and modern periods, when Tuvalu and Kiribati were jointly administered as the British Gilbert and Ellice Islands colony. As these samples were collected from Tuvalu nationals living outside Tuvalu, their particular island of ancestral origin within the nation of Tuvalu was not recorded[20]. For these reasons, we opted to divide Tuvalu into a group we call "Tuvalu Micronesian" and one we call "Tuvalu Polynesian" in subsequent analyses.

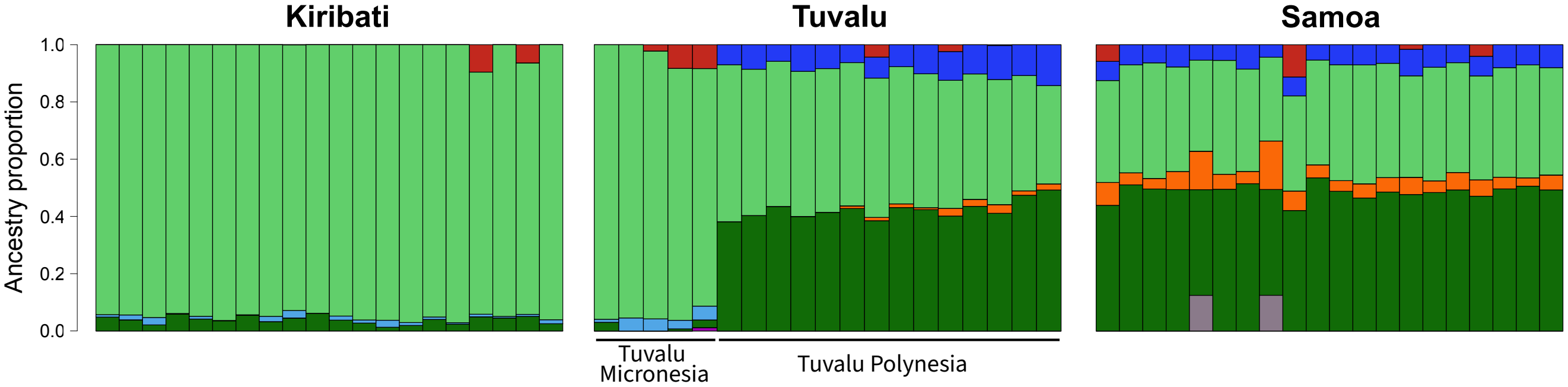

**Supplementary Figure S7.** ADMIXTURE results for K=9 clusters for Kiribati, Tuvalu, and Samoa. Clusters are colored similarly to Figure 1B.

## 6 Principal component analysis of all present-day individuals

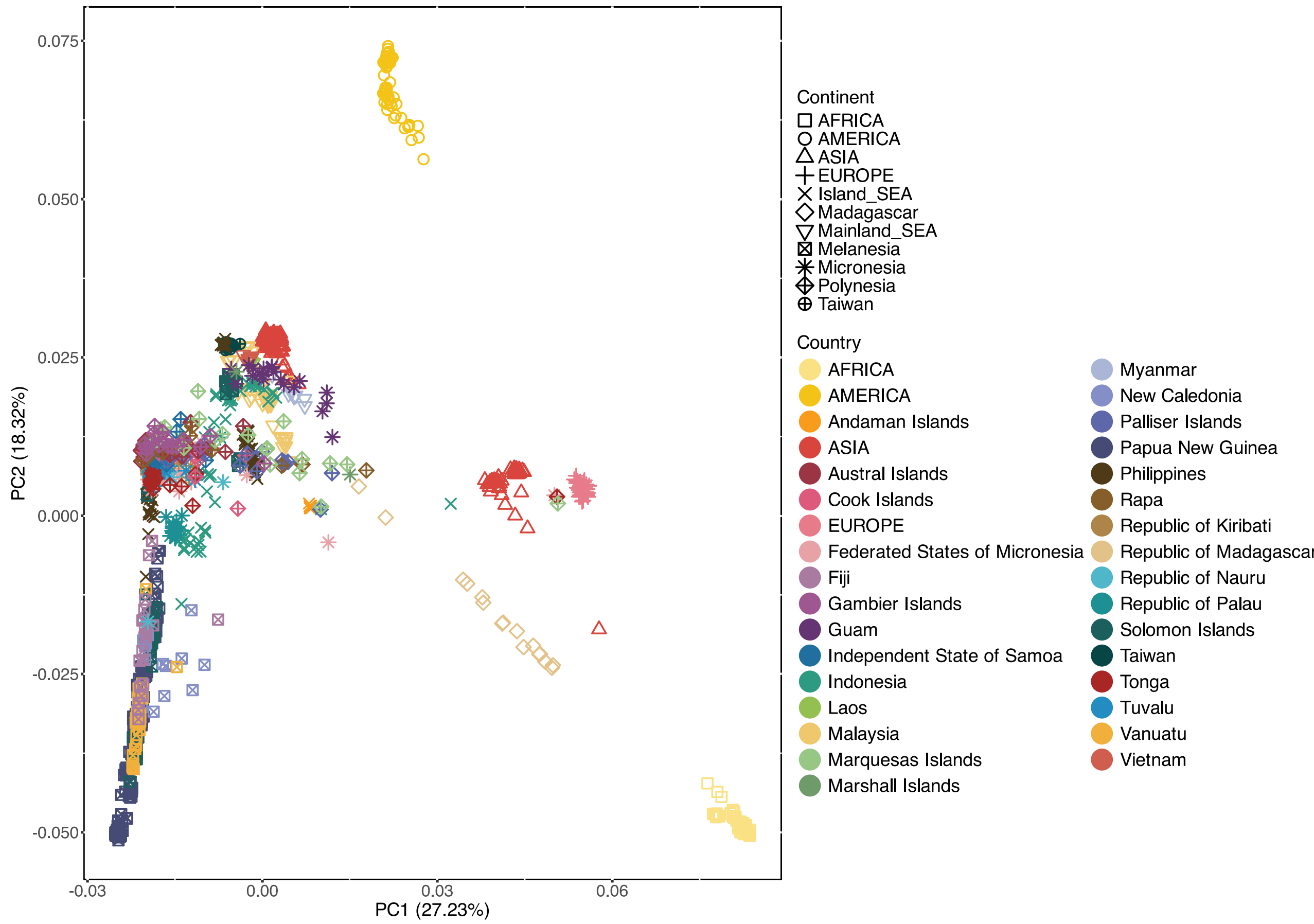

**Supplementary Figure S8.** **A.** Principal component analysis plot of all individuals from OGVP together with selected HGDP populations. These continental reference populations are written in capital letters.

## 7 Ancestry-specific neighbor-joining tree

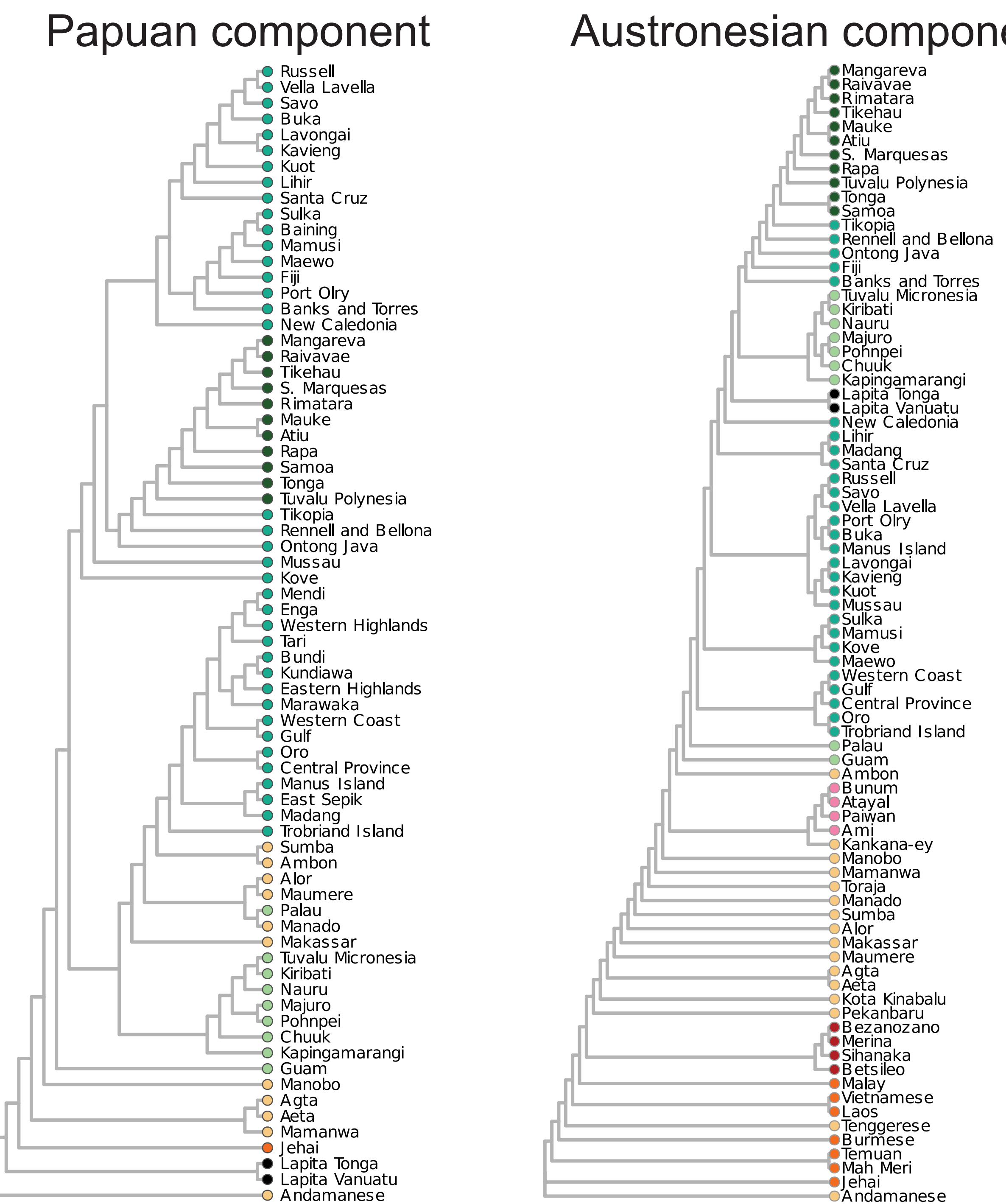

**Supplementary Figure S9.** Ancestry-specific neighbor-joining trees considering only Papuan-like (left) and Austronesian-like (right) chromosomal segments. These trees are identical to those in Figure 2C but with all populations annotated. Populations are colored by region as in Figure 1A.

## 8 Runs of Homozygosity (ROH)

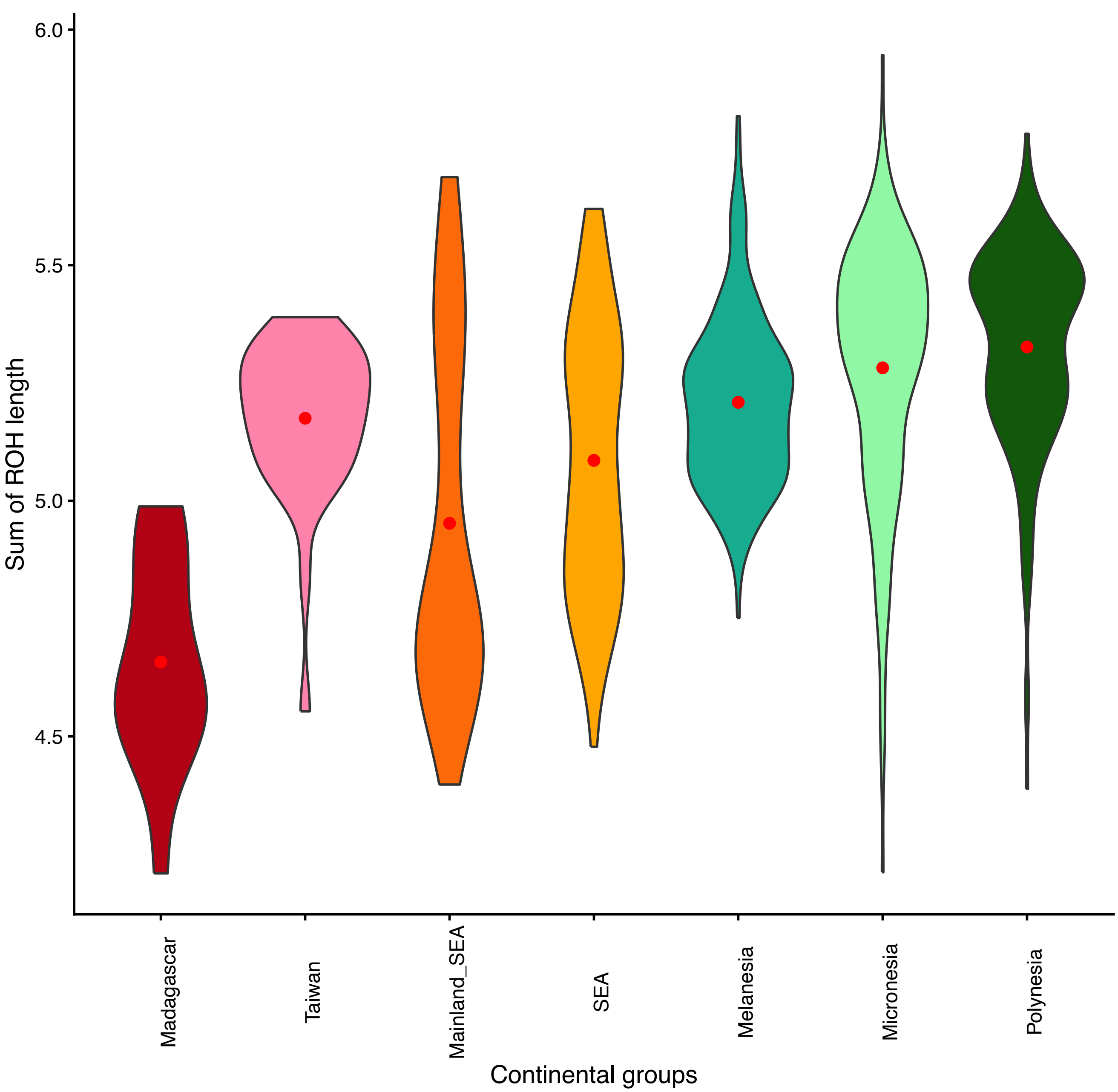

**Supplementary Figure S10.** Distribution of sums of ROH length across different groups.

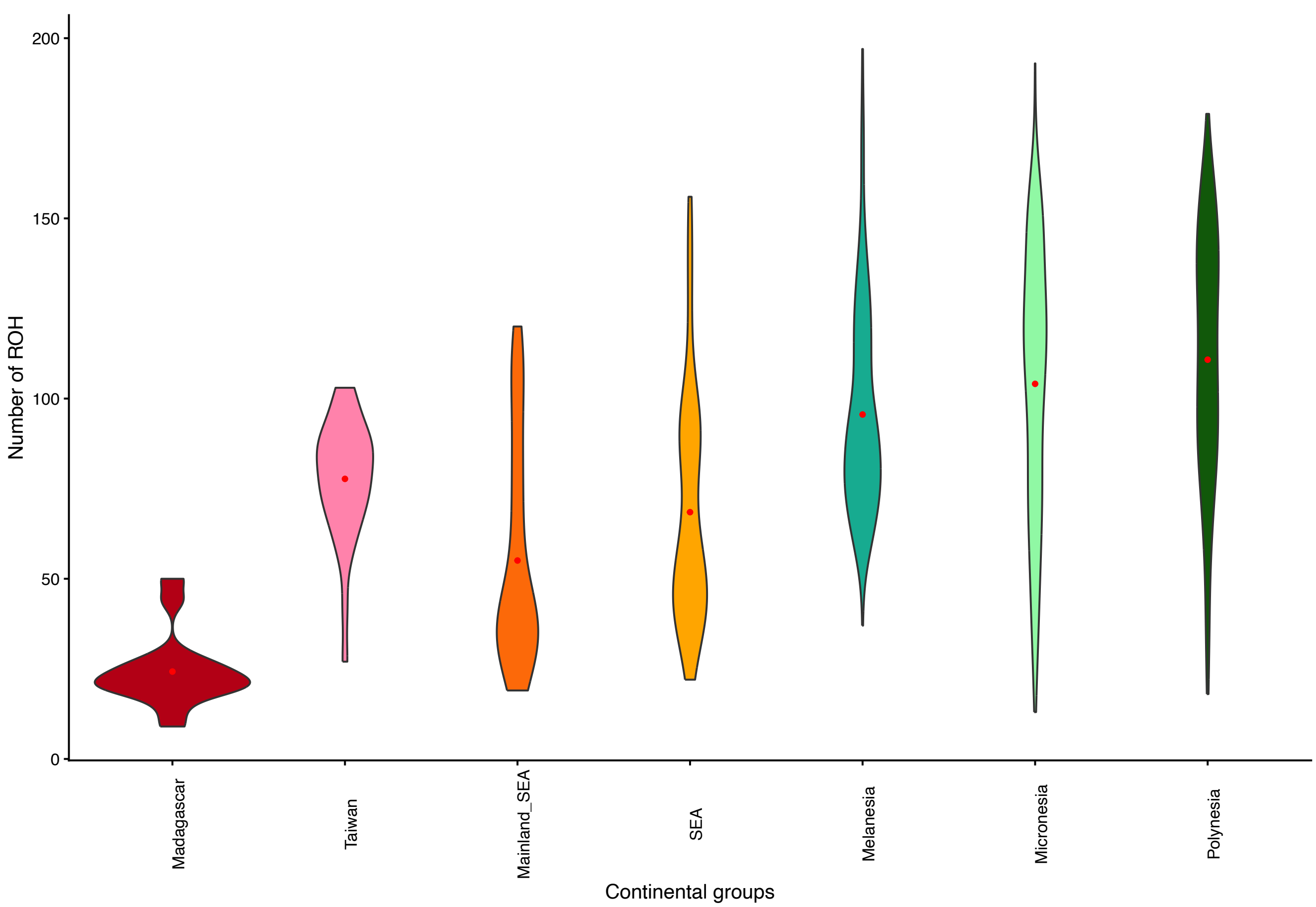

**Supplementary Figure S11.** Distribution of numbers of ROH across different groups.

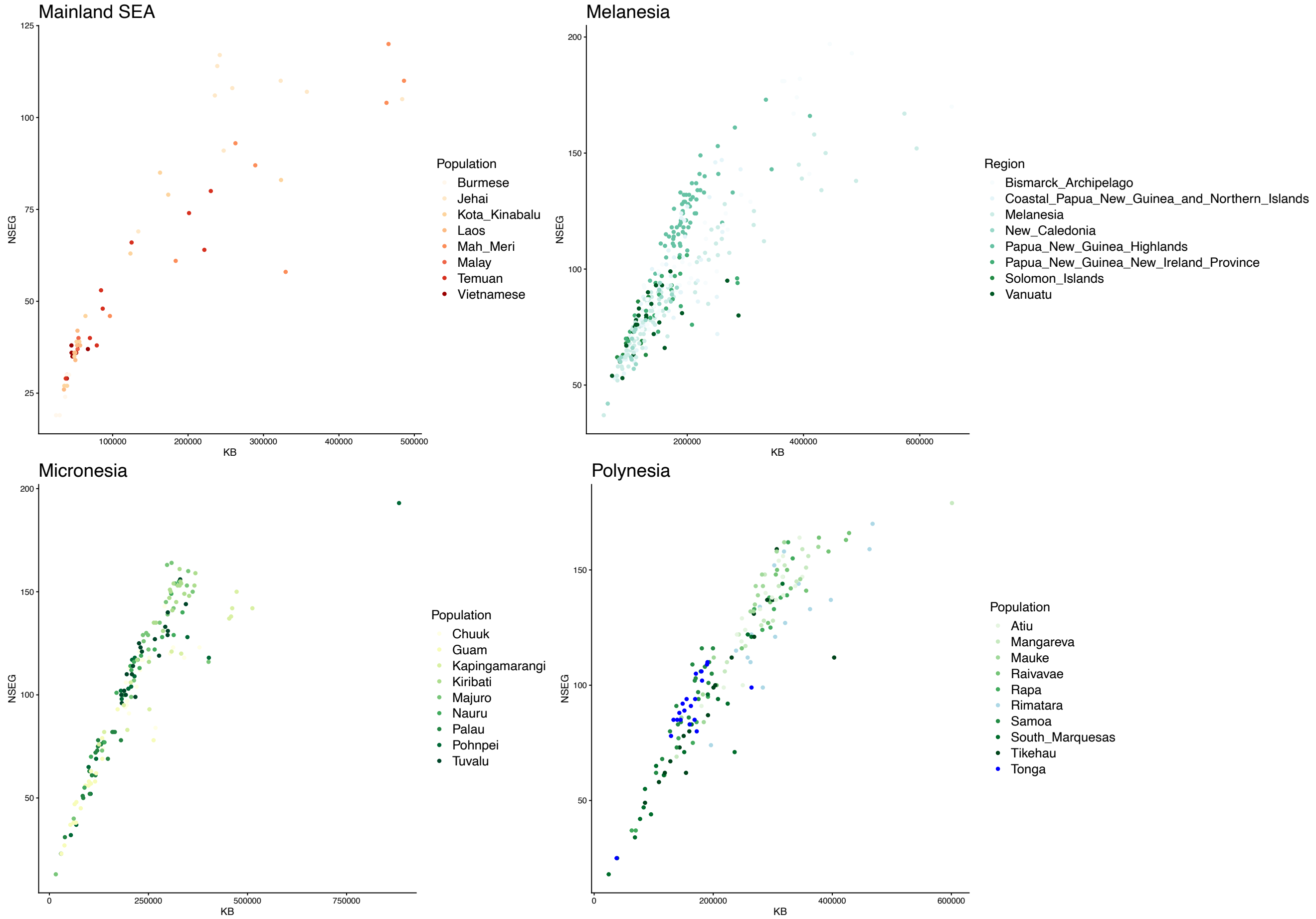

**Supplementary Figure S12.** Comparison of relation of ROH lengths and the number of ROH segments for populations from different groups.

## 9 Outgroup F3 and Polynesian Outliers

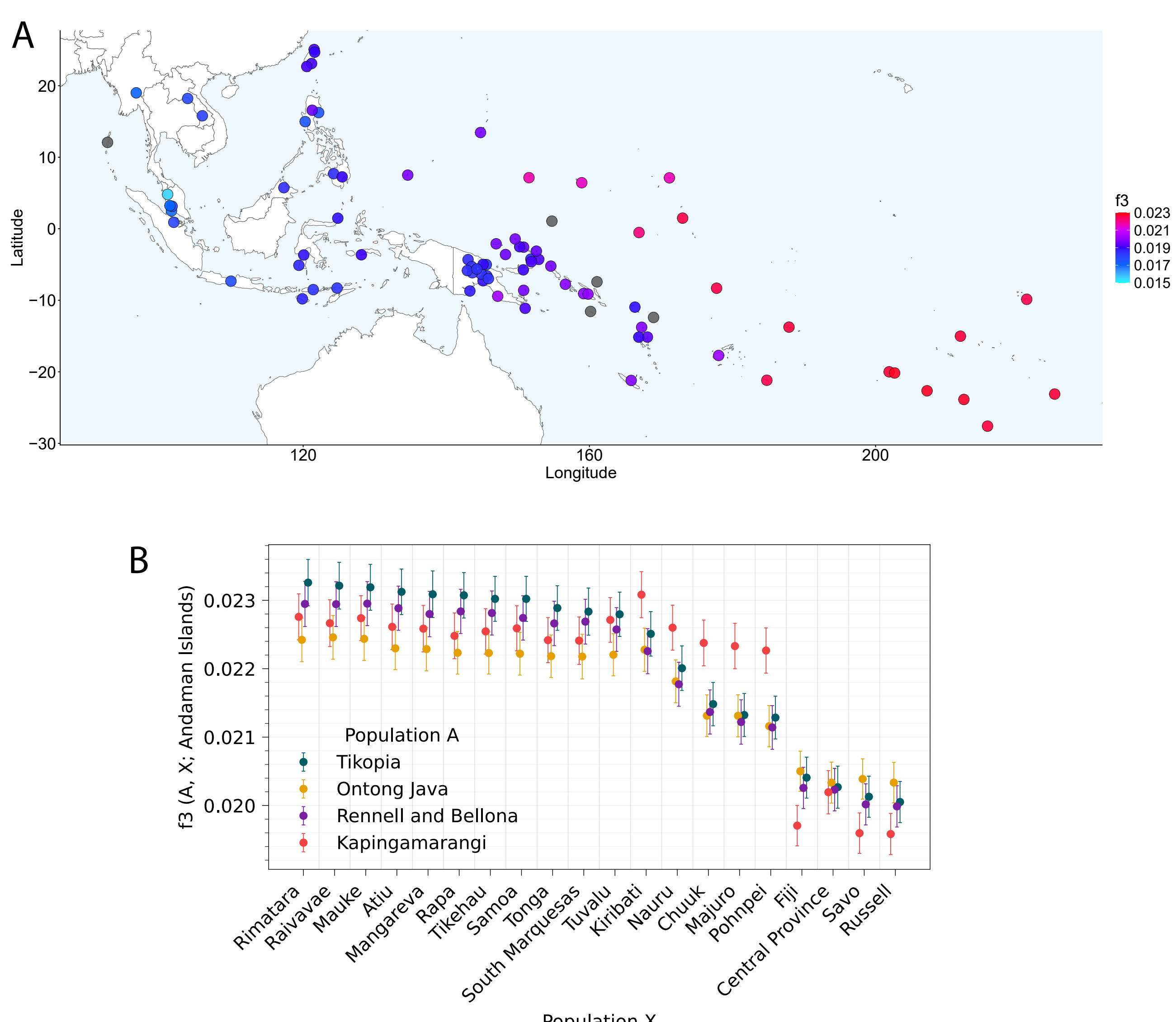

**Supplementary Figure S13.** **A.** Map representing outgroup F3(Polynesian Outliers, X; Andaman Islands) where the Polynesian Outliers are an average value for Tikopia, Ontong Java, Rennell and Bellona, and Kapingamarangi, and X is each of the other populations displayed as dots from blue to pink. The Andaman Islands outgroup are indicated by gray. **B.** Scatterplot showing the top f3 values in **A** for each Polynesian Outlier population separately with error bars determined by jackknife.

**Supplementary Table S1.** Populations used in this study

| Population | Location | Region | Country | Group |
|---|---|---|---|---|
| Betsileo | Madagascar | Madagascar | Republic of Madagascar | Madagascar |
| Bezanozano | Madagascar | Madagascar | Republic of Madagascar | Madagascar |
| Merina | Madagascar | Madagascar | Republic of Madagascar | Madagascar |
| Sihanaka | Madagascar | Madagascar | Republic of Madagascar | Madagascar |
| Ami | - | Taiwan | Taiwan | Taiwan |
| Atayal | - | Taiwan | Taiwan | Taiwan |
| Bunum | - | Taiwan | Taiwan | Taiwan |
| Paiwan | - | Taiwan | Taiwan | Taiwan |
| Burmese | - | Mainland South East Asia | Myanmar | Mainland South East Asia |
| Laos | - | Mainland South East Asia | Laos | Mainland South East Asia |
| Vietnamese | - | Mainland South East Asia | Vietnam | Mainland South East Asia |
| Malay | - | Peninsular South East Asia | Malaysia | Peninsular South East Asia |
| Jehai | - | Peninsular South East Asia | Malaysia | Peninsular South East Asia |
| Mah Meri | - | Peninsular South East Asia | Malaysia | Mainland South East Asia |
| Temuan | - | Peninsular South East Asia | Malaysia | Mainland South East Asia |
| Andamanese | Andaman Islands | Island South East Asia | Andaman Islands | Island South East Asia |
| Aeta | Aglao, Zambales, Luzon | Island South East Asia | Philippines | Island South East Asia |
| Aeta | Sta Juliana, Tarlac, Luzon | Island South East Asia | Philippines | Island South East Asia |
| Agta | Cozo, Aurora, Luzon | Island South East Asia | Philippines | Island South East Asia |
| Agta | Umiray, Aurora, Luzon | Island South East Asia | Philippines | Island South East Asia |
| Kankana-ey | Luzon | Island South East Asia | Philippines | Island South East Asia |
| Mamanwa | Mindanao | Island South East Asia | Philippines | Island South East Asia |
| Manobo | Mindanao | Island South East Asia | Philippines | Island South East Asia |
| Kota Kinabalu | Kota Kinabalu,<br>Sabah, Malaysia | Island South East Asia | Malaysia | Island South East Asia |
| Alor | Alor | Island South East Asia | Indonesia | Island South East Asia |
| Ambon | Ambon | Island South East Asia | Indonesia | Island South East Asia |
| Makassar | South Sulawesi | Island South East Asia | Indonesia | Island South East Asia |
| Manado | North Sulawesi | Island South East Asia | Indonesia | Island South East Asia |
| Maumere | Flores | Island South East Asia | Indonesia | Island South East Asia |
| Pekanbaru | Riau, Sumatra | Island South East Asia | Indonesia | Island outh East Asia |

**Table S1 continued from previous page**

| Population | Location | Region | Country | Group |
|---|---|---|---|---|
| Sumba | Sumba | Island South East Asia | Indonesia | Island South East Asia |
| Tenggerese | Java | Island South East Asia | Indonesia | Island South East Asia |
| Toraja | South Sulawesi | Island South East Asia | Indonesia | Island South East Asia |
| Baining | New Britain, Bismarck Archipelago | Bismarck Archipelago | Papua New Guinea | Melanesia |
| Kove | Bismarck Archipelago | Bismarck Archipelago | Papua New Guinea | Melanesia |
| Lavongai | North, Bismarck Archipelago, New Ireland Province | Bismarck Archipelago | Papua New Guinea | Melanesia |
| Lavongai | South-West, Bismarck Archipelago, New Ireland Province | Bismarck Archipelago | Papua New Guinea | Melanesia |
| Mamusi | New Britain, Bismarck Archipelago | Bismarck Archipelago | Papua New Guinea | Melanesia |
| Manus Island | Manus Province, Bismarck Archipelago | Bismarck Archipelago | Papua New Guinea | Melanesia |
| Mussau | Mussau Island, St. Matthias Islands, Bismarck Archipelago | Bismarck Archipelago | Papua New Guinea | Melanesia |
| Sulka | New Britain, Bismarck Archipelago | Bismarck Archipelago | Papua New Guinea | Melanesia |
| Central Province | - | Coastal Papua New Guinea and Northern Islands | Papua New Guinea | Melanesia |
| East Sepik (Non Austronesian) | East Sepik Province | Coastal Papua New Guinea and Northern Islands | Papua New Guinea | Melanesia |
| Gulf (Non Austronesian) | - | Coastal Papua New Guinea and Northern Islands | Papua New Guinea | Melanesia |
| Madang (Non Austronesian) | Madang Province | Coastal Papua New Guinea and Northern Islands | Papua New Guinea | Melanesia |
| Madang (Austronesian) | Madang Province | Coastal Papua New Guinea and Northern Islands | Papua New Guinea | Melanesia |
| Oro | Oro Province | Coastal Papua New Guinea and Northern Islands | Papua New Guinea | Melanesia |
| Trobriand Islands | Milne Bay Province | Coastal Papua New Guinea and Northern Islands | Papua New Guinea | Melanesia |
| Western Coast | - | Coastal Papua New Guinea and Northern Islands | Papua New Guinea | Melanesia |
| Fiji | - | Melanesia | Fiji | Melanesia |
| Ontong Java | Malaita Province | Melanesia | Solomon Islands | Melanesia |
| Rennell and Bellona | Province of Solomon Islands | Melanesia | Solomon Islands | Melanesia |
| Russell | Pavuvu Island, Russell Islands, Central Province (Lavukaleve speakers) | Melanesia | Solomon Islands | Melanesia |

**Table S1 continued from previous page**

| Population | Location | Region | Country | Group |
|---|---|---|---|---|
| Santa Cruz | Santa Cruz Islands, Temotu Province | Melanesia | Solomon Islands | Melanesia |
| Savo | Savo Island, Central Province | Melanesia | Solomon Islands | Melanesia |
| Tikopia | Temotu Province | Melanesia | Solomon Islands | Melanesia |
| Vella Lavella | Western Province | Melanesia | Solomon Islands | Melanesia |
| New Caledonia | New Caledonia | New Caledonia | New Caledonia | Melanesia |
| Bundi | South Madang Province, Eastern Highlands | Papua New Guinea Highlands | Papua New Guinea | Melanesia |
| Eastern Highlands | Eastern Highlands | Papua New Guinea Highlands | Papua New Guinea | Melanesia |
| Enga | Highlands | Papua New Guinea Highlands | Papua New Guinea | Melanesia |
| Kundiawa | Chimbu Province | Papua New Guinea Highlands | Papua New Guinea | Melanesia |
| Marawaka | Eastern Highlands | Papua New Guinea Highlands | Papua New Guinea | Melanesia |
| Mendi | Central Southern Highlands | Papua New Guinea Highlands | Papua New Guinea | Melanesia |
| Tari | Southern Highlands | Papua New Guinea Highlands | Papua New Guinea | Melanesia |
| Western Highlands | Western Highlands | Papua New Guinea Highlands | Papua New Guinea | Melanesia |
| Kavieng | New Ireland Province | Papua New Guinea New Ireland Province | Papua New Guinea | Melanesia |
| Kuot | New Ireland Province | Papua New Guinea New Ireland Province | Papua New Guinea | Melanesia |
| Lihir | Lihir Island, Bismarck Archipelago | Papua New Guinea New Ireland Province | Papua New Guinea | Melanesia |
| Buka | Buka Island, Boungainville Region | Solomon Islands | Papua New Guinea | Melanesia |
| Banks and Torres | Banks Islands and Torres Islands, Torba, Vanuatu | Vanuatu | Vanuatu | Melanesia |
| Maewo | Maewo Island, Penama Province, Vanuatu | Vanuatu | Vanuatu | Melanesia |
| Port Olry | Espiritu Santo, Vanuatu | Vanuatu | Vanuatu | Melanesia |
| Chuuk | Chuuk State, Federated States of Micronesia | Federated States of Micronesia | Federated States of Micronesia | Micronesia |
| Kapingamarangi | Pohnpei State, Federated States of Micronesia | Federated States of Micronesia | Federated States of Micronesia | Micronesia |
| Pohnpei | Pohnpei State, Federated States of Micronesia | Federated States of Micronesia | Federated States of Micronesia | Micronesia |
| Guam | Guam | Micronesia | Guam | Micronesia |
| Kiribati | Republic of Kiribati | Micronesia | Republic of Kiribati | Micronesia |
| Majuro | Marshall Islands | Micronesia | Marshall Islands | Micronesia |
| Nauru | Republic of Nauru | Micronesia | Republic of Nauru | Micronesia |
| Palau | Republic of Palau | Republic of Palau | Republic of Palau | Micronesia |

**Table S1 continued from previous page**

| Population | Location | Region | Country | Group |
|---|---|---|---|---|
| Tuvalu | Tuvalu | Tuvalu | Tuvalu | Polynesia |
| Atiu | Southern Islands of the Cook Islands | Cook Islands | Cook Islands | Polynesia |
| Mauke | Cook Islands | Cook Islands | Cook Islands | Polynesia |
| Mangareva | Gambier Islands, French Polynesia | French Polynesia | Gambier Islands | Polynesia |
| Raivavae | Austral Islands, French Polynesia | French Polynesia | Austral Islands | Polynesia |
| Rapa | Bass Islands, French Polynesia | French Polynesia | Rapa | Polynesia |
| Rimatara | Austral Islands, French Polynesia | French Polynesia | Austral Islands | Polynesia |
| South Marquesas | Marquesas Islands | French Polynesia | Marquesas Islands | Polynesia |
| Tikehau | Palliser Islands<br>Tuamotu Archipelago, French Polynesia | French Polynesia | Palliser Islands | Polynesia |
| Samoa | Independent State of Samoa | Polynesia | Independent State of Samoa | Polynesia |
| Tonga | Kingdom of Tonga | Polynesia | Tonga | Polynesia |

**Supplementary Table S2.** Ancient and archaic individuals used in this study

| Sample | Location | Country | Group | Mean Date BP (dates) | Latitude | Longitude | Reference(s) |
|---|---|---|---|---|---|---|---|
| I5951 | Efate Island, Vanuatu, Teouma Lapita site | Vanuatu | Melanesia | 2820 (2920–2720) | -17.7860 | 168.3700 | 21 |
| I1368 | Efate Island, Vanuatu, Teouma Lapita site | Vanuatu | Melanesia | 2870 (2990-2740) | -17.7860 | 168.3700 | 22 |
| I1369 | Efate Island, Vanuatu, Teouma Lapita site | Vanuatu | Melanesia | 2885 (3000-2750) | -17.7860 | 168.3700 | 22 |
| I1370 | Efate Island, Vanuatu, Teouma Lapita site | Vanuatu | Melanesia | 2945 (3110-2780) | -17.7860 | 168.3700 | 22 |
| MAL006 | Malakula | Vanuatu | Melanesia | 2495 (2690-2320) | -16.073881 | 167.447786 | 23 |
| TON001 | Tongatapu | Tonga | Polynesia | 2475 (2670-2320) | -21.177119 | -175.109078 | 23 |
| TON002 | Tongatapu | Tonga | Polynesia | 2475 (2690-2350) | -21.177119 | -175.109078 | 23 |
| CP30 | Talasiu Site, Fanga ’Uta Lagoon | Tonga | Polynesia | 2510 (2680-2340) | -21.14518581979385 | -175.2044433513414 | 22 |
| Altai Neanderthal | Denisova Cave,Altai Mountains | Russia | Asia | 50,300 $\pm$ 2,200 years | 51.409 | 84.689 | 24 |
| Altai Denisova | Denisova Cave,Altai Mountains | Russia | Asia | 74,000 - 82,000 years | 51.409 | 84.689 | 25 |

**Supplementary Table S3.** Population groups in OGVP used for the analysis of variants of biomedical relevance.

| Group id | Group name | Number of samples |
|---|---|---|
| 1 | Fiji | 20 |
| 2 | Guam | 20 |
| 3 | Kiribati | 20 |
| 4 | Majuro | 20 |
| 5 | New Caledonia | 20 |
| 6 | Samoa | 20 |
| 7 | Tonga | 20 |
| 8 | Chuuk | 19 |
| 9 | Palau | 19 |
| 10 | South Marquesas | 19 |
| 11 | Tuvalu | 19 |
| 12 | Atiu | 18 |
| 13 | Nauru | 18 |
| 14 | Tikehau | 18 |
| 15 | Aeta | 17 |
| 16 | Mangareva | 17 |
| 17 | Pohnpei | 17 |
| 18 | Agta | 16 |
| 19 | Rimatara | 16 |
| 20 | Mauke | 15 |
| 21 | Rapa | 15 |
| 22 | Raivavae | 14 |
| 23 | Kapingamarangi | 13 |
| 24 | Lavongai, Kuot | 19 |
| 25 | Kavieng, Lihir | 17 |
| 26 | Kove, Mussau, Manus Island | 19 |
| 27 | East New Britain | 24 |
| 28 | Southern Papua New Guinea | 32 |
| 29 | Madang, East Sepik | 17 |
| 30 | Northern Vanuatu | 28 |
| 31 | West Papuan Highlands | 32 |
| 32 | East Papuan Highlands | 30 |
| 33 | Taiwan | 32 |
| 34 | Andaman Islands | 8 |
| 35 | Kankana-ey | 10 |
| 36 | Sunda arc | 26 |
| 37 | Sulawesi and Ambon Islands | 30 |
| 38 | Myanmar, Laos, Vietnam | 24 |
| 39 | Orang Asli | 26 |
| 40 | Malay, Kota Kinabalu | 15 |
| 41 | Malagasy | 16 |

### SM-Tables - Variants of biomedical relevance

- **SM-clinvar-drugResponse-with-gnomad.tsv** Clinvar drug response variant frequencies in OGVP and gnomAD.
- **SM-clinvar-likelyPathogenic-with-gnomad.tsv** Clinvar likely pathogenic variant frequencies in OGVP and gnomAD.
- **SM-clinvar-pathogenic-with-gnomad.tsv** Clinvar pathogenic variant frequencies in OGVP and gnomAD.
- **SM-pharmacogenomic-with-gnomad.tsv** Pharmacogenomic variant frequencies in OGVP and gnomAD.